\documentclass[a4paper,11pt]{article}

    \usepackage[breakable]{tcolorbox}
    \usepackage{parskip} 

    \usepackage{graphicx}
    
    \usepackage{caption}
    \DeclareCaptionFormat{nocaption}{}
    \usepackage{float}
    \usepackage{xcolor} 
    \usepackage{enumerate} 
    \usepackage{geometry} 
    \usepackage{amsmath} 
    \usepackage{amssymb} 
    \usepackage{textcomp} 
    \AtBeginDocument{%
        \def\PYZsq{\textquotesingle}
    }
    \usepackage{upquote} 
    \usepackage{eurosym} 

    \usepackage{iftex}
    \ifPDFTeX
        \usepackage[T1]{fontenc}
        \IfFileExists{alphabeta.sty}{
              \usepackage{alphabeta}
          }{
              \usepackage[mathletters]{ucs}
              \usepackage[utf8x]{inputenc}
          }
    \else
        \usepackage{fontspec}
        \usepackage{unicode-math}
    \fi

    \usepackage{fancyvrb} 
    \usepackage{grffile} 
    \makeatletter 
    \@ifpackagelater{grffile}{2019/11/01}
    {
    }
    {
      \def\Gread@@xetex#1{%
        \IfFileExists{"\Gin@base".bb}%
        {\Gread@eps{\Gin@base.bb}}%
        {\Gread@@xetex@aux#1}%
      }
    }
    \makeatother
    \usepackage[Export]{adjustbox} 
    \adjustboxset{max size={0.9\linewidth}{0.9\paperheight}}

    \usepackage{hyperref}
    \usepackage{titling}
    \usepackage{longtable} 
    \usepackage{booktabs}  
    \usepackage{array}     
    \usepackage{calc}      
    \usepackage[inline]{enumitem} 
    \usepackage[normalem]{ulem} 
    \usepackage{mathrsfs}

    \definecolor{urlcolor}{rgb}{0,.145,.698}
    \definecolor{linkcolor}{rgb}{.71,0.21,0.01}
    \definecolor{citecolor}{rgb}{.12,.54,.11}

    \definecolor{ansi-black}{HTML}{3E424D}
    \definecolor{ansi-black-intense}{HTML}{282C36}
    \definecolor{ansi-red}{HTML}{E75C58}
    \definecolor{ansi-red-intense}{HTML}{B22B31}
    \definecolor{ansi-green}{HTML}{00A250}
    \definecolor{ansi-green-intense}{HTML}{007427}
    \definecolor{ansi-yellow}{HTML}{DDB62B}
    \definecolor{ansi-yellow-intense}{HTML}{B27D12}
    \definecolor{ansi-blue}{HTML}{208FFB}
    \definecolor{ansi-blue-intense}{HTML}{0065CA}
    \definecolor{ansi-magenta}{HTML}{D160C4}
    \definecolor{ansi-magenta-intense}{HTML}{A03196}
    \definecolor{ansi-cyan}{HTML}{60C6C8}
    \definecolor{ansi-cyan-intense}{HTML}{258F8F}
    \definecolor{ansi-white}{HTML}{C5C1B4}
    \definecolor{ansi-white-intense}{HTML}{A1A6B2}
    \definecolor{ansi-default-inverse-fg}{HTML}{FFFFFF}
    \definecolor{ansi-default-inverse-bg}{HTML}{000000}

    \definecolor{outerrorbackground}{HTML}{FFDFDF}

    \DefineVerbatimEnvironment{Highlighting}{Verbatim}{commandchars=\\\{\}}

    \let\Oldtex\TeX
    \let\Oldlatex\LaTeX
    \renewcommand{\TeX}{\textrm{\Oldtex}}
    \renewcommand{\LaTeX}{\textrm{\Oldlatex}}
    \title{Tutorial: Drinfeld modules in SageMath}
    \author{David Ayotte, Xavier Caruso, Antoine Leudière, Joseph Musleh} 
    \date{\small Try online at \url{https://caruso.perso.math.cnrs.fr/notebook/drinfeld-modules}}

\makeatletter
\def\PY@reset{\let\PY@it=\relax \let\PY@bf=\relax%
    \let\PY@ul=\relax \let\PY@tc=\relax%
    \let\PY@bc=\relax \let\PY@ff=\relax}
\def\PY@tok#1{\csname PY@tok@#1\endcsname}
\def\PY@toks#1+{\ifx\relax#1\empty\else%
    \PY@tok{#1}\expandafter\PY@toks\fi}
\def\PY@do#1{\PY@bc{\PY@tc{\PY@ul{%
    \PY@it{\PY@bf{\PY@ff{#1}}}}}}}
\def\PY#1#2{\PY@reset\PY@toks#1+\relax+\PY@do{#2}}

\@namedef{PY@tok@w}{\def\PY@tc##1{\textcolor[rgb]{0.73,0.73,0.73}{##1}}}
\@namedef{PY@tok@c}{\let\PY@it=\textit\def\PY@tc##1{\textcolor[rgb]{0.24,0.48,0.48}{##1}}}
\@namedef{PY@tok@cp}{\def\PY@tc##1{\textcolor[rgb]{0.61,0.40,0.00}{##1}}}
\@namedef{PY@tok@k}{\let\PY@bf=\textbf\def\PY@tc##1{\textcolor[rgb]{0.00,0.50,0.00}{##1}}}
\@namedef{PY@tok@kp}{\def\PY@tc##1{\textcolor[rgb]{0.00,0.50,0.00}{##1}}}
\@namedef{PY@tok@kt}{\def\PY@tc##1{\textcolor[rgb]{0.69,0.00,0.25}{##1}}}
\@namedef{PY@tok@o}{\def\PY@tc##1{\textcolor[rgb]{0.40,0.40,0.40}{##1}}}
\@namedef{PY@tok@ow}{\let\PY@bf=\textbf\def\PY@tc##1{\textcolor[rgb]{0.67,0.13,1.00}{##1}}}
\@namedef{PY@tok@nb}{\def\PY@tc##1{\textcolor[rgb]{0.00,0.50,0.00}{##1}}}
\@namedef{PY@tok@nf}{\def\PY@tc##1{\textcolor[rgb]{0.00,0.00,1.00}{##1}}}
\@namedef{PY@tok@nc}{\let\PY@bf=\textbf\def\PY@tc##1{\textcolor[rgb]{0.00,0.00,1.00}{##1}}}
\@namedef{PY@tok@nn}{\let\PY@bf=\textbf\def\PY@tc##1{\textcolor[rgb]{0.00,0.00,1.00}{##1}}}
\@namedef{PY@tok@ne}{\let\PY@bf=\textbf\def\PY@tc##1{\textcolor[rgb]{0.80,0.25,0.22}{##1}}}
\@namedef{PY@tok@nv}{\def\PY@tc##1{\textcolor[rgb]{0.10,0.09,0.49}{##1}}}
\@namedef{PY@tok@no}{\def\PY@tc##1{\textcolor[rgb]{0.53,0.00,0.00}{##1}}}
\@namedef{PY@tok@nl}{\def\PY@tc##1{\textcolor[rgb]{0.46,0.46,0.00}{##1}}}
\@namedef{PY@tok@ni}{\let\PY@bf=\textbf\def\PY@tc##1{\textcolor[rgb]{0.44,0.44,0.44}{##1}}}
\@namedef{PY@tok@na}{\def\PY@tc##1{\textcolor[rgb]{0.41,0.47,0.13}{##1}}}
\@namedef{PY@tok@nt}{\let\PY@bf=\textbf\def\PY@tc##1{\textcolor[rgb]{0.00,0.50,0.00}{##1}}}
\@namedef{PY@tok@nd}{\def\PY@tc##1{\textcolor[rgb]{0.67,0.13,1.00}{##1}}}
\@namedef{PY@tok@s}{\def\PY@tc##1{\textcolor[rgb]{0.73,0.13,0.13}{##1}}}
\@namedef{PY@tok@sd}{\let\PY@it=\textit\def\PY@tc##1{\textcolor[rgb]{0.73,0.13,0.13}{##1}}}
\@namedef{PY@tok@si}{\let\PY@bf=\textbf\def\PY@tc##1{\textcolor[rgb]{0.64,0.35,0.47}{##1}}}
\@namedef{PY@tok@se}{\let\PY@bf=\textbf\def\PY@tc##1{\textcolor[rgb]{0.67,0.36,0.12}{##1}}}
\@namedef{PY@tok@sr}{\def\PY@tc##1{\textcolor[rgb]{0.64,0.35,0.47}{##1}}}
\@namedef{PY@tok@ss}{\def\PY@tc##1{\textcolor[rgb]{0.10,0.09,0.49}{##1}}}
\@namedef{PY@tok@sx}{\def\PY@tc##1{\textcolor[rgb]{0.00,0.50,0.00}{##1}}}
\@namedef{PY@tok@m}{\def\PY@tc##1{\textcolor[rgb]{0.40,0.40,0.40}{##1}}}
\@namedef{PY@tok@gh}{\let\PY@bf=\textbf\def\PY@tc##1{\textcolor[rgb]{0.00,0.00,0.50}{##1}}}
\@namedef{PY@tok@gu}{\let\PY@bf=\textbf\def\PY@tc##1{\textcolor[rgb]{0.50,0.00,0.50}{##1}}}
\@namedef{PY@tok@gd}{\def\PY@tc##1{\textcolor[rgb]{0.63,0.00,0.00}{##1}}}
\@namedef{PY@tok@gi}{\def\PY@tc##1{\textcolor[rgb]{0.00,0.52,0.00}{##1}}}
\@namedef{PY@tok@gr}{\def\PY@tc##1{\textcolor[rgb]{0.89,0.00,0.00}{##1}}}
\@namedef{PY@tok@ge}{\let\PY@it=\textit}
\@namedef{PY@tok@gs}{\let\PY@bf=\textbf}
\@namedef{PY@tok@gp}{\let\PY@bf=\textbf\def\PY@tc##1{\textcolor[rgb]{0.00,0.00,0.50}{##1}}}
\@namedef{PY@tok@go}{\def\PY@tc##1{\textcolor[rgb]{0.44,0.44,0.44}{##1}}}
\@namedef{PY@tok@gt}{\def\PY@tc##1{\textcolor[rgb]{0.00,0.27,0.87}{##1}}}
\@namedef{PY@tok@err}{\def\PY@bc##1{{\setlength{\fboxsep}{\string -\fboxrule}\fcolorbox[rgb]{1.00,0.00,0.00}{1,1,1}{\strut ##1}}}}
\@namedef{PY@tok@kc}{\let\PY@bf=\textbf\def\PY@tc##1{\textcolor[rgb]{0.00,0.50,0.00}{##1}}}
\@namedef{PY@tok@kd}{\let\PY@bf=\textbf\def\PY@tc##1{\textcolor[rgb]{0.00,0.50,0.00}{##1}}}
\@namedef{PY@tok@kn}{\let\PY@bf=\textbf\def\PY@tc##1{\textcolor[rgb]{0.00,0.50,0.00}{##1}}}
\@namedef{PY@tok@kr}{\let\PY@bf=\textbf\def\PY@tc##1{\textcolor[rgb]{0.00,0.50,0.00}{##1}}}
\@namedef{PY@tok@bp}{\def\PY@tc##1{\textcolor[rgb]{0.00,0.50,0.00}{##1}}}
\@namedef{PY@tok@fm}{\def\PY@tc##1{\textcolor[rgb]{0.00,0.00,1.00}{##1}}}
\@namedef{PY@tok@vc}{\def\PY@tc##1{\textcolor[rgb]{0.10,0.09,0.49}{##1}}}
\@namedef{PY@tok@vg}{\def\PY@tc##1{\textcolor[rgb]{0.10,0.09,0.49}{##1}}}
\@namedef{PY@tok@vi}{\def\PY@tc##1{\textcolor[rgb]{0.10,0.09,0.49}{##1}}}
\@namedef{PY@tok@vm}{\def\PY@tc##1{\textcolor[rgb]{0.10,0.09,0.49}{##1}}}
\@namedef{PY@tok@sa}{\def\PY@tc##1{\textcolor[rgb]{0.73,0.13,0.13}{##1}}}
\@namedef{PY@tok@sb}{\def\PY@tc##1{\textcolor[rgb]{0.73,0.13,0.13}{##1}}}
\@namedef{PY@tok@sc}{\def\PY@tc##1{\textcolor[rgb]{0.73,0.13,0.13}{##1}}}
\@namedef{PY@tok@dl}{\def\PY@tc##1{\textcolor[rgb]{0.73,0.13,0.13}{##1}}}
\@namedef{PY@tok@s2}{\def\PY@tc##1{\textcolor[rgb]{0.73,0.13,0.13}{##1}}}
\@namedef{PY@tok@sh}{\def\PY@tc##1{\textcolor[rgb]{0.73,0.13,0.13}{##1}}}
\@namedef{PY@tok@s1}{\def\PY@tc##1{\textcolor[rgb]{0.73,0.13,0.13}{##1}}}
\@namedef{PY@tok@mb}{\def\PY@tc##1{\textcolor[rgb]{0.40,0.40,0.40}{##1}}}
\@namedef{PY@tok@mf}{\def\PY@tc##1{\textcolor[rgb]{0.40,0.40,0.40}{##1}}}
\@namedef{PY@tok@mh}{\def\PY@tc##1{\textcolor[rgb]{0.40,0.40,0.40}{##1}}}
\@namedef{PY@tok@mi}{\def\PY@tc##1{\textcolor[rgb]{0.40,0.40,0.40}{##1}}}
\@namedef{PY@tok@il}{\def\PY@tc##1{\textcolor[rgb]{0.40,0.40,0.40}{##1}}}
\@namedef{PY@tok@mo}{\def\PY@tc##1{\textcolor[rgb]{0.40,0.40,0.40}{##1}}}
\@namedef{PY@tok@ch}{\let\PY@it=\textit\def\PY@tc##1{\textcolor[rgb]{0.24,0.48,0.48}{##1}}}
\@namedef{PY@tok@cm}{\let\PY@it=\textit\def\PY@tc##1{\textcolor[rgb]{0.24,0.48,0.48}{##1}}}
\@namedef{PY@tok@cpf}{\let\PY@it=\textit\def\PY@tc##1{\textcolor[rgb]{0.24,0.48,0.48}{##1}}}
\@namedef{PY@tok@c1}{\let\PY@it=\textit\def\PY@tc##1{\textcolor[rgb]{0.24,0.48,0.48}{##1}}}
\@namedef{PY@tok@cs}{\let\PY@it=\textit\def\PY@tc##1{\textcolor[rgb]{0.24,0.48,0.48}{##1}}}

\def\PYZus{\char`\_}
\def\PYZob{\char`\{}
\def\PYZcb{\char`\}}
\def\PYZca{\char`\^}

\def\PYZlt{\char`\<}
\def\PYZgt{\char`\>}
\def\PYZsh{\char`\#}
\def\PYZpc{\char`\%}

\def\PYZhy{\char`\-}
\def\PYZsq{\char`\'}
\def\PYZdq{\char`\"}

\makeatother

    \makeatletter
        \newbox\Wrappedcontinuationbox
        \newbox\Wrappedvisiblespacebox
        \newcommand*\Wrappedvisiblespace {\textcolor{red}{\textvisiblespace}}
        \newcommand*\Wrappedcontinuationsymbol {\textcolor{red}{\llap{\tiny$\m@th\hookrightarrow$}}}
        \newcommand*\Wrappedcontinuationindent {3ex }
        \newcommand*\Wrappedafterbreak {\kern\Wrappedcontinuationindent\copy\Wrappedcontinuationbox}
        \newcommand*\Wrappedbreaksatspecials {%
            \def\PYGZus{\discretionary{\char`\_}{\Wrappedafterbreak}{\char`\_}}%
            \def\PYGZob{\discretionary{}{\Wrappedafterbreak\char`\{}{\char`\{}}%
            \def\PYGZcb{\discretionary{\char`\}}{\Wrappedafterbreak}{\char`\}}}%
            \def\PYGZca{\discretionary{\char`\^}{\Wrappedafterbreak}{\char`\^}}%
            \def\PYGZam{\discretionary{\char`\&}{\Wrappedafterbreak}{\char`\&}}%
            \def\PYGZlt{\discretionary{}{\Wrappedafterbreak\char`\<}{\char`\<}}%
            \def\PYGZgt{\discretionary{\char`\>}{\Wrappedafterbreak}{\char`\>}}%
            \def\PYGZsh{\discretionary{}{\Wrappedafterbreak\char`\#}{\char`\#}}%
            \def\PYGZpc{\discretionary{}{\Wrappedafterbreak\char`\%}{\char`\%}}%
            \def\PYGZdl{\discretionary{}{\Wrappedafterbreak\char`\$}{\char`\$}}%
            \def\PYGZhy{\discretionary{\char`\-}{\Wrappedafterbreak}{\char`\-}}%
            \def\PYGZsq{\discretionary{}{\Wrappedafterbreak\textquotesingle}{\textquotesingle}}%
            \def\PYGZdq{\discretionary{}{\Wrappedafterbreak\char`\"}{\char`\"}}%
            \def\PYGZti{\discretionary{\char`\~}{\Wrappedafterbreak}{\char`\~}}%
        }
        \newcommand*\Wrappedbreaksatpunct {%
            \lccode`\~`\.\lowercase{\def~}{\discretionary{\hbox{\char`\.}}{\Wrappedafterbreak}{\hbox{\char`\.}}}%
            \lccode`\~`\,\lowercase{\def~}{\discretionary{\hbox{\char`\,}}{\Wrappedafterbreak}{\hbox{\char`\,}}}%
            \lccode`\~`\;\lowercase{\def~}{\discretionary{\hbox{\char`\;}}{\Wrappedafterbreak}{\hbox{\char`\;}}}%
            \lccode`\~`\:\lowercase{\def~}{\discretionary{\hbox{\char`\:}}{\Wrappedafterbreak}{\hbox{\char`\:}}}%
            \lccode`\~`\?\lowercase{\def~}{\discretionary{\hbox{\char`\?}}{\Wrappedafterbreak}{\hbox{\char`\?}}}%
            \lccode`\~`\!\lowercase{\def~}{\discretionary{\hbox{\char`\!}}{\Wrappedafterbreak}{\hbox{\char`\!}}}%
            \lccode`\~`\/\lowercase{\def~}{\discretionary{\hbox{\char`\/}}{\Wrappedafterbreak}{\hbox{\char`\/}}}%
            \catcode`\.\active
            \catcode`\,\active
            \catcode`\;\active
            \catcode`\:\active
            \catcode`\?\active
            \catcode`\!\active
            \catcode`\/\active
            \lccode`\~`\~
        }
    \makeatother

    \let\OriginalVerbatim=\Verbatim
    \makeatletter
    \renewcommand{\Verbatim}[1][1]{%
        \sbox\Wrappedcontinuationbox {\Wrappedcontinuationsymbol}%
        \sbox\Wrappedvisiblespacebox {\FV@SetupFont\Wrappedvisiblespace}%
        \def\FancyVerbFormatLine ##1{\hsize\linewidth
            \vtop{\raggedright\hyphenpenalty\z@\exhyphenpenalty\z@
                \doublehyphendemerits\z@\finalhyphendemerits\z@
                \strut ##1\strut}%
        }%
        \def\FV@Space {%
            \nobreak\hskip\z@ plus\fontdimen3\font minus\fontdimen4\font
            \discretionary{\copy\Wrappedvisiblespacebox}{\Wrappedafterbreak}
            {\kern\fontdimen2\font}%
        }%

        \Wrappedbreaksatspecials
        \OriginalVerbatim[#1,codes*=\Wrappedbreaksatpunct]%
    }
    \makeatother

    \definecolor{incolor}{HTML}{303F9F}
    \definecolor{outcolor}{HTML}{D84315}
    \definecolor{cellborder}{HTML}{CFCFCF}
    \definecolor{cellbackground}{HTML}{F7F7F7}

    \makeatletter
    \newcommand{\boxspacing}{\kern\kvtcb@left@rule\kern\kvtcb@boxsep}
    \makeatother
    \newcommand{\prompt}[4]{
        {\ttfamily\llap{{\color{#2}[#3]:\hspace{3pt}#4}}\vspace{-\baselineskip}}
    }

    \hypersetup{
      breaklinks=true,  
      colorlinks=true,
      urlcolor=urlcolor,
      linkcolor=linkcolor,
      citecolor=citecolor,
      }
    
\begin{document}

\maketitle

\setcounter{tocdepth}{1}
\tableofcontents
\setcounter{tocdepth}{3}

\bigskip

\noindent\hfill\hrulefill\hfill\null

\bigskip

    \hypertarget{drinfeld-modules-construction-and-basic-properties}{%
\section{\texorpdfstring{Construction and basic properties}{Construction and basic properties}}\label{drinfeld-modules-construction-and-basic-properties}}

    \hypertarget{definitions}{%
\subsection{Definitions}\label{definitions}}

Let \(\mathbb{F}_q\) denote a finite field with \(q\) elements. We
consider a \(\mathbb{F}_q\)-algebra \(A\) and an \(A\)-algebra \(K\). We
denote by \(\gamma : A \to K\) the defining morphism of \(K\).

Let also \(K\{\tau\}\) denote the ring of \emph{skew} polynomials over
\(K\) in the variable \(\tau\), that is the ring of usual polynomials
with multiplication twisted by the rule

\[\tau a = a^q \tau, \quad \forall a \in K.\]

By definition, a \(A\)-Drinfeld module is a ring homomorphism
\(\phi : A \to K\{\tau\}\) such that 
\begin{itemize}
\item for all \(a \in A\), the constant
coefficient of \(\phi(a)\) is \(\gamma(a)\),
\item for all \(a \in A\),
\(a \not\in \mathbb{F}_q\), the skew polynomial \(\phi(a)\) has positive
degree.
\end{itemize}

We often write \(\phi_a\) instead of \(\phi(a)\).

In what follows, we will only consider the case where
\(A = \mathbb{F}_q[T]\) (our implementation is limited to this setting
so far) and will simply say \emph{Drinfeld module} instead of
\(\mathbb{F}_q[T]\)-Drinfeld module. We remark that a Drinfeld module
\(\phi\) is entirely defined by the datum of \(\phi_T\).

    \hypertarget{construction-in-sagemath}{%
\subsection{Construction in SageMath}\label{construction-in-sagemath}}

We first define \(\mathbb{F}_q\), \(A\) and \(K\).

    \begin{tcolorbox}[breakable, size=fbox, boxrule=1pt, pad at break*=1mm,colback=cellbackground, colframe=cellborder]
\prompt{In}{incolor}{1}{\boxspacing}
\begin{Verbatim}[commandchars=\\\{\}]
\PY{n}{Fq} \PY{o}{=} \PY{n}{GF}\PY{p}{(}\PY{l+m+mi}{5}\PY{p}{)}
\PY{n}{A}\PY{o}{.}\PY{o}{\PYZlt{}}\PY{n}{T}\PY{o}{\PYZgt{}} \PY{o}{=} \PY{n}{Fq}\PY{p}{[}\PY{p}{]}
\PY{n}{K}\PY{o}{.}\PY{o}{\PYZlt{}}\PY{n}{z}\PY{o}{\PYZgt{}} \PY{o}{=} \PY{n}{Fq}\PY{o}{.}\PY{n}{extension}\PY{p}{(}\PY{l+m+mi}{3}\PY{p}{)}
\end{Verbatim}
\end{tcolorbox}

    We can now create a Drinfeld module using the constructor
\texttt{DrinfeldModule}:

    \begin{tcolorbox}[breakable, size=fbox, boxrule=1pt, pad at break*=1mm,colback=cellbackground, colframe=cellborder]
\prompt{In}{incolor}{2}{\boxspacing}
\begin{Verbatim}[commandchars=\\\{\}]
\PY{n}{phi} \PY{o}{=} \PY{n}{DrinfeldModule}\PY{p}{(}\PY{n}{A}\PY{p}{,} \PY{p}{[}\PY{n}{z}\PY{p}{,} \PY{l+m+mi}{0}\PY{p}{,} \PY{l+m+mi}{1}\PY{p}{,} \PY{l+m+mi}{1}\PY{p}{]}\PY{p}{)}
\PY{n}{phi}
\end{Verbatim}
\end{tcolorbox}

            \begin{tcolorbox}[breakable, size=fbox, boxrule=.5pt, pad at break*=1mm, opacityfill=0]
\prompt{Out}{outcolor}{2}{\boxspacing}
\begin{Verbatim}[commandchars=\\\{\}]
Drinfeld module defined by T |--> t\^{}3 + t\^{}2 + z
\end{Verbatim}
\end{tcolorbox}
        
    The first argument of the constructor is the base ring \texttt{A}. The
second argument can be either:
\begin{itemize}
\item the skew polynomial \(\phi_T\), or
\item the list of coefficients of \(\phi_T\) (as in the example above).
\end{itemize}

In both cases, one can recover the underlying skew polynomial ring using
the method \texttt{ore\_polring}:

    \begin{tcolorbox}[breakable, size=fbox, boxrule=1pt, pad at break*=1mm,colback=cellbackground, colframe=cellborder]
\prompt{In}{incolor}{3}{\boxspacing}
\begin{Verbatim}[commandchars=\\\{\}]
\PY{n}{phi}\PY{o}{.}\PY{n}{ore\PYZus{}polring}\PY{p}{(}\PY{p}{)}
\end{Verbatim}
\end{tcolorbox}

            \begin{tcolorbox}[breakable, size=fbox, boxrule=.5pt, pad at break*=1mm, opacityfill=0]
\prompt{Out}{outcolor}{3}{\boxspacing}
\begin{Verbatim}[commandchars=\\\{\}]
Ore Polynomial Ring in t over Finite Field in z of size 5\^{}3 over its base
twisted by Frob
\end{Verbatim}
\end{tcolorbox}
        
    The image of a polynomial \(a \in A\) can be obtained by simply calling
\texttt{phi(a)}:

    \begin{tcolorbox}[breakable, size=fbox, boxrule=1pt, pad at break*=1mm,colback=cellbackground, colframe=cellborder]
\prompt{In}{incolor}{4}{\boxspacing}
\begin{Verbatim}[commandchars=\\\{\}]
\PY{n}{phi}\PY{p}{(}\PY{n}{T}\PY{p}{)}
\end{Verbatim}
\end{tcolorbox}

            \begin{tcolorbox}[breakable, size=fbox, boxrule=.5pt, pad at break*=1mm, opacityfill=0]
\prompt{Out}{outcolor}{4}{\boxspacing}
\begin{Verbatim}[commandchars=\\\{\}]
t\^{}3 + t\^{}2 + z
\end{Verbatim}
\end{tcolorbox}
        
    \begin{tcolorbox}[breakable, size=fbox, boxrule=1pt, pad at break*=1mm,colback=cellbackground, colframe=cellborder]
\prompt{In}{incolor}{5}{\boxspacing}
\begin{Verbatim}[commandchars=\\\{\}]
\PY{n}{phi}\PY{p}{(}\PY{n}{T}\PY{o}{\PYZca{}}\PY{l+m+mi}{2} \PY{o}{+} \PY{l+m+mi}{1}\PY{p}{)}
\end{Verbatim}
\end{tcolorbox}

            \begin{tcolorbox}[breakable, size=fbox, boxrule=.5pt, pad at break*=1mm, opacityfill=0]
\prompt{Out}{outcolor}{5}{\boxspacing}
\begin{Verbatim}[commandchars=\\\{\}]
t\^{}6 + 2*t\^{}5 + t\^{}4 + 2*z*t\^{}3 + (3*z\^{}2 + z + 1)*t\^{}2 + z\^{}2 + 1
\end{Verbatim}
\end{tcolorbox}
        
    \begin{tcolorbox}[breakable, size=fbox, boxrule=1pt, pad at break*=1mm,colback=cellbackground, colframe=cellborder]
\prompt{In}{incolor}{6}{\boxspacing}
\begin{Verbatim}[commandchars=\\\{\}]
\PY{n}{phi}\PY{p}{(}\PY{n}{T}\PY{o}{\PYZca{}}\PY{l+m+mi}{2} \PY{o}{+} \PY{l+m+mi}{1}\PY{p}{)} \PY{o}{==} \PY{n}{phi}\PY{p}{(}\PY{n}{T}\PY{p}{)}\PY{o}{\PYZca{}}\PY{l+m+mi}{2} \PY{o}{+} \PY{l+m+mi}{1}   \PY{c+c1}{\PYZsh{} basic check: phi is a ring homomorphism}
\end{Verbatim}
\end{tcolorbox}

            \begin{tcolorbox}[breakable, size=fbox, boxrule=.5pt, pad at break*=1mm, opacityfill=0]
\prompt{Out}{outcolor}{6}{\boxspacing}
\begin{Verbatim}[commandchars=\\\{\}]
True
\end{Verbatim}
\end{tcolorbox}
        
    \hypertarget{basic-properties}{%
\subsection{Basic properties}\label{basic-properties}}

Many standard invariants attached to Drinfeld modules can be computed.

\hypertarget{characteristic}{%
\subsubsection{Characteristic}\label{characteristic}}

The \emph{characteristic} of a Drinfeld module \(\phi\) is, by
definition, (a generator of) the kernel of \(\gamma\).

    \begin{tcolorbox}[breakable, size=fbox, boxrule=1pt, pad at break*=1mm,colback=cellbackground, colframe=cellborder]
\prompt{In}{incolor}{7}{\boxspacing}
\begin{Verbatim}[commandchars=\\\{\}]
\PY{n}{phi}\PY{o}{.}\PY{n}{characteristic}\PY{p}{(}\PY{p}{)}
\end{Verbatim}
\end{tcolorbox}

            \begin{tcolorbox}[breakable, size=fbox, boxrule=.5pt, pad at break*=1mm, opacityfill=0]
\prompt{Out}{outcolor}{7}{\boxspacing}
\begin{Verbatim}[commandchars=\\\{\}]
T\^{}3 + 3*T + 3
\end{Verbatim}
\end{tcolorbox}
        
    \hypertarget{rank}{%
\subsubsection{Rank}\label{rank}}

The \emph{rank} of a Drinfeld module \(\phi\) is, by definition, the
degree of the skew polynomial \(\phi_T\). More generally, it satisfies
the relation:

\[\forall a \in \mathbb{F}_q[T], \quad \deg \phi_a = \text{rank}(\phi) \cdot \deg a.\]

    \begin{tcolorbox}[breakable, size=fbox, boxrule=1pt, pad at break*=1mm,colback=cellbackground, colframe=cellborder]
\prompt{In}{incolor}{8}{\boxspacing}
\begin{Verbatim}[commandchars=\\\{\}]
\PY{n}{phi}\PY{o}{.}\PY{n}{rank}\PY{p}{(}\PY{p}{)}
\end{Verbatim}
\end{tcolorbox}

            \begin{tcolorbox}[breakable, size=fbox, boxrule=.5pt, pad at break*=1mm, opacityfill=0]
\prompt{Out}{outcolor}{8}{\boxspacing}
\begin{Verbatim}[commandchars=\\\{\}]
3
\end{Verbatim}
\end{tcolorbox}
        
    \begin{tcolorbox}[breakable, size=fbox, boxrule=1pt, pad at break*=1mm,colback=cellbackground, colframe=cellborder]
\prompt{In}{incolor}{9}{\boxspacing}
\begin{Verbatim}[commandchars=\\\{\}]
\PY{n}{a} \PY{o}{=} \PY{n}{T}\PY{o}{\PYZca{}}\PY{l+m+mi}{2} \PY{o}{+} \PY{l+m+mi}{1}
\PY{n}{phi}\PY{p}{(}\PY{n}{a}\PY{p}{)}\PY{o}{.}\PY{n}{degree}\PY{p}{(}\PY{p}{)} \PY{o}{==} \PY{n}{phi}\PY{o}{.}\PY{n}{rank}\PY{p}{(}\PY{p}{)} \PY{o}{*} \PY{n}{a}\PY{o}{.}\PY{n}{degree}\PY{p}{(}\PY{p}{)}
\end{Verbatim}
\end{tcolorbox}

            \begin{tcolorbox}[breakable, size=fbox, boxrule=.5pt, pad at break*=1mm, opacityfill=0]
\prompt{Out}{outcolor}{9}{\boxspacing}
\begin{Verbatim}[commandchars=\\\{\}]
True
\end{Verbatim}
\end{tcolorbox}
        
    \hypertarget{height}{%
\subsubsection{Height}\label{height}}

Let \(\mathfrak p\) be the characteristic of \(\phi\). The height of
\(\phi\) is, by definition, the quotient of the \(\tau\)-valuation of
\(\phi_{\mathfrak p}\) be the degree of \(\mathfrak p\); it is always a
positive integer.

\textbf{Remark}: The height is an important invariant because it is
related to the rank of the \(\mathfrak p\)-torsion points of the
Drinfeld module.

    \begin{tcolorbox}[breakable, size=fbox, boxrule=1pt, pad at break*=1mm,colback=cellbackground, colframe=cellborder]
\prompt{In}{incolor}{10}{\boxspacing}
\begin{Verbatim}[commandchars=\\\{\}]
\PY{n}{phi}\PY{o}{.}\PY{n}{height}\PY{p}{(}\PY{p}{)}
\end{Verbatim}
\end{tcolorbox}

            \begin{tcolorbox}[breakable, size=fbox, boxrule=.5pt, pad at break*=1mm, opacityfill=0]
\prompt{Out}{outcolor}{10}{\boxspacing}
\begin{Verbatim}[commandchars=\\\{\}]
1
\end{Verbatim}
\end{tcolorbox}
        
    Having height 1 is the generic case. However, Drinfeld modules with
higher heights also exist. Here is an example in rank 2:

    \begin{tcolorbox}[breakable, size=fbox, boxrule=1pt, pad at break*=1mm,colback=cellbackground, colframe=cellborder]
\prompt{In}{incolor}{11}{\boxspacing}
\begin{Verbatim}[commandchars=\\\{\}]
\PY{n}{psi} \PY{o}{=} \PY{n}{DrinfeldModule}\PY{p}{(}\PY{n}{A}\PY{p}{,} \PY{p}{[}\PY{n}{z}\PY{p}{,} \PY{l+m+mi}{0}\PY{p}{,} \PY{l+m+mi}{1}\PY{p}{]}\PY{p}{)}
\PY{n}{psi}\PY{o}{.}\PY{n}{height}\PY{p}{(}\PY{p}{)}
\end{Verbatim}
\end{tcolorbox}

            \begin{tcolorbox}[breakable, size=fbox, boxrule=.5pt, pad at break*=1mm, opacityfill=0]
\prompt{Out}{outcolor}{11}{\boxspacing}
\begin{Verbatim}[commandchars=\\\{\}]
2
\end{Verbatim}
\end{tcolorbox}
        
    \hypertarget{categories-of-drinfeld-modules}{%
\subsection{Categories of Drinfeld
modules}\label{categories-of-drinfeld-modules}}

When a Drinfeld module is creates, SageMath automatically creates at the
same time the category in which it lives. This category remembers the
base rings \(A\) and \(K\) and the defining morphism
\(\gamma : A \to K\).

    \begin{tcolorbox}[breakable, size=fbox, boxrule=1pt, pad at break*=1mm,colback=cellbackground, colframe=cellborder]
\prompt{In}{incolor}{12}{\boxspacing}
\begin{Verbatim}[commandchars=\\\{\}]
\PY{n}{C} \PY{o}{=} \PY{n}{phi}\PY{o}{.}\PY{n}{category}\PY{p}{(}\PY{p}{)}
\PY{n}{C}
\end{Verbatim}
\end{tcolorbox}

            \begin{tcolorbox}[breakable, size=fbox, boxrule=.5pt, pad at break*=1mm, opacityfill=0]
\prompt{Out}{outcolor}{12}{\boxspacing}
\begin{Verbatim}[commandchars=\\\{\}]
Category of Drinfeld modules over Finite Field in z of size 5\^{}3 over its base
\end{Verbatim}
\end{tcolorbox}
        
    Having this category as an actual object in SageMath is important for
dealing with morphisms between Drinfeld modules (as we shall see later
on). Besides, several important invariants are defined at the level of
the category:

    \begin{tcolorbox}[breakable, size=fbox, boxrule=1pt, pad at break*=1mm,colback=cellbackground, colframe=cellborder]
\prompt{In}{incolor}{13}{\boxspacing}
\begin{Verbatim}[commandchars=\\\{\}]
\PY{n}{C}\PY{o}{.}\PY{n}{base}\PY{p}{(}\PY{p}{)}            \PY{c+c1}{\PYZsh{} The field K, vue as an algebra over A }
\end{Verbatim}
\end{tcolorbox}

            \begin{tcolorbox}[breakable, size=fbox, boxrule=.5pt, pad at break*=1mm, opacityfill=0]
\prompt{Out}{outcolor}{13}{\boxspacing}
\begin{Verbatim}[commandchars=\\\{\}]
Finite Field in z of size 5\^{}3 over its base
\end{Verbatim}
\end{tcolorbox}
        
    \begin{tcolorbox}[breakable, size=fbox, boxrule=1pt, pad at break*=1mm,colback=cellbackground, colframe=cellborder]
\prompt{In}{incolor}{14}{\boxspacing}
\begin{Verbatim}[commandchars=\\\{\}]
\PY{n}{C}\PY{o}{.}\PY{n}{base\PYZus{}morphism}\PY{p}{(}\PY{p}{)}   \PY{c+c1}{\PYZsh{} The defining morphism gamma}
                    \PY{c+c1}{\PYZsh{} also accessible by C.base().defining\PYZus{}morphism()}
\end{Verbatim}
\end{tcolorbox}

            \begin{tcolorbox}[breakable, size=fbox, boxrule=.5pt, pad at break*=1mm, opacityfill=0]
\prompt{Out}{outcolor}{14}{\boxspacing}
\begin{Verbatim}[commandchars=\\\{\}]
Ring morphism:
  From: Univariate Polynomial Ring in T over Finite Field of size 5
  To:   Finite Field in z of size 5\^{}3 over its base
  Defn: T |--> z
\end{Verbatim}
\end{tcolorbox}
        
    \begin{tcolorbox}[breakable, size=fbox, boxrule=1pt, pad at break*=1mm,colback=cellbackground, colframe=cellborder]
\prompt{In}{incolor}{15}{\boxspacing}
\begin{Verbatim}[commandchars=\\\{\}]
\PY{n}{C}\PY{o}{.}\PY{n}{ore\PYZus{}polring}\PY{p}{(}\PY{p}{)}     \PY{c+c1}{\PYZsh{} The ring K\PYZob{}t\PYZcb{}}
\end{Verbatim}
\end{tcolorbox}

            \begin{tcolorbox}[breakable, size=fbox, boxrule=.5pt, pad at break*=1mm, opacityfill=0]
\prompt{Out}{outcolor}{15}{\boxspacing}
\begin{Verbatim}[commandchars=\\\{\}]
Ore Polynomial Ring in t over Finite Field in z of size 5\^{}3 over its base
twisted by Frob
\end{Verbatim}
\end{tcolorbox}
        
    \begin{tcolorbox}[breakable, size=fbox, boxrule=1pt, pad at break*=1mm,colback=cellbackground, colframe=cellborder]
\prompt{In}{incolor}{16}{\boxspacing}
\begin{Verbatim}[commandchars=\\\{\}]
\PY{n}{C}\PY{o}{.}\PY{n}{characteristic}\PY{p}{(}\PY{p}{)}
\end{Verbatim}
\end{tcolorbox}

            \begin{tcolorbox}[breakable, size=fbox, boxrule=.5pt, pad at break*=1mm, opacityfill=0]
\prompt{Out}{outcolor}{16}{\boxspacing}
\begin{Verbatim}[commandchars=\\\{\}]
T\^{}3 + 3*T + 3
\end{Verbatim}
\end{tcolorbox}
        

\newpage

    \hypertarget{drinfeld-modules-morphisms-and-isogenies}{%
\section{\texorpdfstring{Morphisms and
isogenies}{Morphisms and isogenies}}\label{drinfeld-modules-morphisms-and-isogenies}}

    \hypertarget{definitions}{%
\subsection{Definitions}\label{definitions}}

Let \(\phi, \psi : \mathbb{F}_q[T] \to K\{\tau\}\) be two Drinfeld
modules in the same category.

By definition, a \emph{morphism} \(\phi \to \psi\) is the datum of
\(u \in K\{\tau\}\) such that \(u \phi_T = \psi_T u\). (This relation
implies more generally that \(u \phi_a = \psi_a u\) for all
\(a \in \mathbb{F}_q[T]\).)

An \emph{isogeny} is, by definition, a nonzero morphism. We observe that
if an isogeny \(\phi \to \psi\) exists, then \(\phi\) and \(\psi\) must
have the same rank.

    \hypertarget{construction}{%
\subsection{Construction}\label{construction}}

    \begin{tcolorbox}[breakable, size=fbox, boxrule=1pt, pad at break*=1mm,colback=cellbackground, colframe=cellborder]
\prompt{In}{incolor}{1}{\boxspacing}
\begin{Verbatim}[commandchars=\\\{\}]
\PY{n}{Fq} \PY{o}{=} \PY{n}{GF}\PY{p}{(}\PY{l+m+mi}{5}\PY{p}{)}
\PY{n}{A}\PY{o}{.}\PY{o}{\PYZlt{}}\PY{n}{T}\PY{o}{\PYZgt{}} \PY{o}{=} \PY{n}{Fq}\PY{p}{[}\PY{p}{]}
\PY{n}{K}\PY{o}{.}\PY{o}{\PYZlt{}}\PY{n}{z}\PY{o}{\PYZgt{}} \PY{o}{=} \PY{n}{Fq}\PY{o}{.}\PY{n}{extension}\PY{p}{(}\PY{l+m+mi}{3}\PY{p}{)}
\PY{n}{phi} \PY{o}{=} \PY{n}{DrinfeldModule}\PY{p}{(}\PY{n}{A}\PY{p}{,} \PY{p}{[}\PY{n}{z}\PY{p}{,} \PY{l+m+mi}{0}\PY{p}{,} \PY{l+m+mi}{1}\PY{p}{,} \PY{n}{z}\PY{p}{]}\PY{p}{)}
\end{Verbatim}
\end{tcolorbox}

    \hypertarget{the-constructor-hom}{%
\subsubsection{\texorpdfstring{The constructor
\texttt{hom}}{The constructor hom}}\label{the-constructor-hom}}

The simplest way to create a morphism is to use the method \texttt{hom},
to which we can directly pass in the defining skew polynomial:

    \begin{tcolorbox}[breakable, size=fbox, boxrule=1pt, pad at break*=1mm,colback=cellbackground, colframe=cellborder]
\prompt{In}{incolor}{2}{\boxspacing}
\begin{Verbatim}[commandchars=\\\{\}]
\PY{n}{t} \PY{o}{=} \PY{n}{phi}\PY{o}{.}\PY{n}{ore\PYZus{}variable}\PY{p}{(}\PY{p}{)}
\PY{n}{f} \PY{o}{=} \PY{n}{phi}\PY{o}{.}\PY{n}{hom}\PY{p}{(}\PY{n}{t} \PY{o}{+} \PY{l+m+mi}{1}\PY{p}{)}
\PY{n}{f}
\end{Verbatim}
\end{tcolorbox}

            \begin{tcolorbox}[breakable, size=fbox, boxrule=.5pt, pad at break*=1mm, opacityfill=0]
\prompt{Out}{outcolor}{2}{\boxspacing}
\begin{Verbatim}[commandchars=\\\{\}]
Drinfeld Module morphism:
  From: Drinfeld module defined by T |--> z*t\^{}3 + t\^{}2 + z
  To:   Drinfeld module defined by T |--> (2*z\^{}2 + 4*z + 4)*t\^{}3 + (3*z\^{}2 + 2*z +
2)*t\^{}2 + (2*z\^{}2 + 3*z + 4)*t + z
  Defn: t + 1
\end{Verbatim}
\end{tcolorbox}
        
    On the example above, we observe that the codomain of the isogeny (which
is not \(\phi\)) was automatically determined:

    \begin{tcolorbox}[breakable, size=fbox, boxrule=1pt, pad at break*=1mm,colback=cellbackground, colframe=cellborder]
\prompt{In}{incolor}{3}{\boxspacing}
\begin{Verbatim}[commandchars=\\\{\}]
\PY{n}{psi} \PY{o}{=} \PY{n}{f}\PY{o}{.}\PY{n}{codomain}\PY{p}{(}\PY{p}{)}
\PY{n}{psi}
\end{Verbatim}
\end{tcolorbox}

            \begin{tcolorbox}[breakable, size=fbox, boxrule=.5pt, pad at break*=1mm, opacityfill=0]
\prompt{Out}{outcolor}{3}{\boxspacing}
\begin{Verbatim}[commandchars=\\\{\}]
Drinfeld module defined by T |--> (2*z\^{}2 + 4*z + 4)*t\^{}3 + (3*z\^{}2 + 2*z + 2)*t\^{}2
+ (2*z\^{}2 + 3*z + 4)*t + z
\end{Verbatim}
\end{tcolorbox}
        
    An important class of endomorphisms of a given Drinfeld module \(\phi\)
are scalar multiplications: they are endomorphisms corresponding to the
skew polynomial \(\phi_a\) for \(a \in \mathbb{F}_q[T]\). Those
endomorphisms can be simply instantiated as follows:

    \begin{tcolorbox}[breakable, size=fbox, boxrule=1pt, pad at break*=1mm,colback=cellbackground, colframe=cellborder]
\prompt{In}{incolor}{4}{\boxspacing}
\begin{Verbatim}[commandchars=\\\{\}]
\PY{n}{g} \PY{o}{=} \PY{n}{phi}\PY{o}{.}\PY{n}{hom}\PY{p}{(}\PY{n}{T}\PY{p}{)}
\PY{n}{g}
\end{Verbatim}
\end{tcolorbox}

            \begin{tcolorbox}[breakable, size=fbox, boxrule=.5pt, pad at break*=1mm, opacityfill=0]
\prompt{Out}{outcolor}{4}{\boxspacing}
\begin{Verbatim}[commandchars=\\\{\}]
Endomorphism of Drinfeld module defined by T |--> z*t\^{}3 + t\^{}2 + z
  Defn: z*t\^{}3 + t\^{}2 + z
\end{Verbatim}
\end{tcolorbox}
        
    \hypertarget{set-of-morphisms-between-two-drinfeld-modules}{%
\subsubsection{Set of morphisms between two Drinfeld
modules}\label{set-of-morphisms-between-two-drinfeld-modules}}

When we know the domain \(\phi\) and the codomain \(\psi\), we can
construct the \emph{homset} \(\text{Hom}(\phi, \psi)\):

    \begin{tcolorbox}[breakable, size=fbox, boxrule=1pt, pad at break*=1mm,colback=cellbackground, colframe=cellborder]
\prompt{In}{incolor}{5}{\boxspacing}
\begin{Verbatim}[commandchars=\\\{\}]
\PY{n}{H} \PY{o}{=} \PY{n}{Hom}\PY{p}{(}\PY{n}{phi}\PY{p}{,} \PY{n}{psi}\PY{p}{)}
\PY{n}{H}
\end{Verbatim}
\end{tcolorbox}

            \begin{tcolorbox}[breakable, size=fbox, boxrule=.5pt, pad at break*=1mm, opacityfill=0]
\prompt{Out}{outcolor}{5}{\boxspacing}
\begin{Verbatim}[commandchars=\\\{\}]
Set of Drinfeld module morphisms from (gen) z*t\^{}3 + t\^{}2 + z to (gen) (2*z\^{}2 +
4*z + 4)*t\^{}3 + (3*z\^{}2 + 2*z + 2)*t\^{}2 + (2*z\^{}2 + 3*z + 4)*t + z
\end{Verbatim}
\end{tcolorbox}
        
    As a shortcut, when \(\phi = \psi\), we can use the \texttt{End}
primitive:

    \begin{tcolorbox}[breakable, size=fbox, boxrule=1pt, pad at break*=1mm,colback=cellbackground, colframe=cellborder]
\prompt{In}{incolor}{6}{\boxspacing}
\begin{Verbatim}[commandchars=\\\{\}]
\PY{n}{E} \PY{o}{=} \PY{n}{End}\PY{p}{(}\PY{n}{phi}\PY{p}{)}
\PY{n}{E}
\end{Verbatim}
\end{tcolorbox}

            \begin{tcolorbox}[breakable, size=fbox, boxrule=.5pt, pad at break*=1mm, opacityfill=0]
\prompt{Out}{outcolor}{6}{\boxspacing}
\begin{Verbatim}[commandchars=\\\{\}]
Set of Drinfeld module morphisms from (gen) z*t\^{}3 + t\^{}2 + z to (gen) z*t\^{}3 + t\^{}2
+ z
\end{Verbatim}
\end{tcolorbox}
        
    Notice that those homsets are the parents in which \texttt{f} and
\texttt{g} (which we created earlier) naturally live:

    \begin{tcolorbox}[breakable, size=fbox, boxrule=1pt, pad at break*=1mm,colback=cellbackground, colframe=cellborder]
\prompt{In}{incolor}{7}{\boxspacing}
\begin{Verbatim}[commandchars=\\\{\}]
\PY{n}{f}\PY{o}{.}\PY{n}{parent}\PY{p}{(}\PY{p}{)} \PY{o+ow}{is} \PY{n}{H}
\end{Verbatim}
\end{tcolorbox}

            \begin{tcolorbox}[breakable, size=fbox, boxrule=.5pt, pad at break*=1mm, opacityfill=0]
\prompt{Out}{outcolor}{7}{\boxspacing}
\begin{Verbatim}[commandchars=\\\{\}]
True
\end{Verbatim}
\end{tcolorbox}
        
    \begin{tcolorbox}[breakable, size=fbox, boxrule=1pt, pad at break*=1mm,colback=cellbackground, colframe=cellborder]
\prompt{In}{incolor}{8}{\boxspacing}
\begin{Verbatim}[commandchars=\\\{\}]
\PY{n}{g}\PY{o}{.}\PY{n}{parent}\PY{p}{(}\PY{p}{)} \PY{o+ow}{is} \PY{n}{E}
\end{Verbatim}
\end{tcolorbox}

            \begin{tcolorbox}[breakable, size=fbox, boxrule=.5pt, pad at break*=1mm, opacityfill=0]
\prompt{Out}{outcolor}{8}{\boxspacing}
\begin{Verbatim}[commandchars=\\\{\}]
True
\end{Verbatim}
\end{tcolorbox}
        
    Once the homset is defined, we can call it to create morphisms in it (as
it is usual in SageMath). This provides an alternative to the
\texttt{hom} constructor to define morphisms. For example:

    \begin{tcolorbox}[breakable, size=fbox, boxrule=1pt, pad at break*=1mm,colback=cellbackground, colframe=cellborder]
\prompt{In}{incolor}{9}{\boxspacing}
\begin{Verbatim}[commandchars=\\\{\}]
\PY{n}{Frob} \PY{o}{=} \PY{n}{E}\PY{p}{(}\PY{n}{t}\PY{o}{\PYZca{}}\PY{l+m+mi}{3}\PY{p}{)}
\PY{n}{Frob}
\end{Verbatim}
\end{tcolorbox}

            \begin{tcolorbox}[breakable, size=fbox, boxrule=.5pt, pad at break*=1mm, opacityfill=0]
\prompt{Out}{outcolor}{9}{\boxspacing}
\begin{Verbatim}[commandchars=\\\{\}]
Endomorphism of Drinfeld module defined by T |--> z*t\^{}3 + t\^{}2 + z
  Defn: t\^{}3
\end{Verbatim}
\end{tcolorbox}
        
    The latter is the Frobenius endomorphism. It is also available
\emph{via} the method \texttt{frobenius\_endomorphism}:

    \begin{tcolorbox}[breakable, size=fbox, boxrule=1pt, pad at break*=1mm,colback=cellbackground, colframe=cellborder]
\prompt{In}{incolor}{10}{\boxspacing}
\begin{Verbatim}[commandchars=\\\{\}]
\PY{n}{phi}\PY{o}{.}\PY{n}{frobenius\PYZus{}endomorphism}\PY{p}{(}\PY{p}{)}
\end{Verbatim}
\end{tcolorbox}

            \begin{tcolorbox}[breakable, size=fbox, boxrule=.5pt, pad at break*=1mm, opacityfill=0]
\prompt{Out}{outcolor}{10}{\boxspacing}
\begin{Verbatim}[commandchars=\\\{\}]
Endomorphism of Drinfeld module defined by T |--> z*t\^{}3 + t\^{}2 + z
  Defn: t\^{}3
\end{Verbatim}
\end{tcolorbox}
        
    \hypertarget{about-the-structure-of-texthomphi-psi-and-textendphi}{%
\subsection{\texorpdfstring{About the structure of
\(\text{Hom}(\phi, \psi)\) and
\(\text{End}(\phi)\)}{About the structure of \textbackslash text\{Hom\}(\textbackslash phi, \textbackslash psi) and \textbackslash text\{End\}(\textbackslash phi)}}\label{about-the-structure-of-texthomphi-psi-and-textendphi}}

First of all, of course, there is a composition law on the homsets: when
the domain of \(f\) is equal to the codomain of \(g\), one can compose
\(f\) and \(g\), producing a new morphism \(f \circ g\). In the Drinfeld
module setting, this operation simply corresponds to the multiplication
of the underlying skew polynomials.

It is available in SageMath using the multiplication operator \texttt{*}
or the exponentiation operator \texttt{\^{}} (or \texttt{**}):

    \begin{tcolorbox}[breakable, size=fbox, boxrule=1pt, pad at break*=1mm,colback=cellbackground, colframe=cellborder]
\prompt{In}{incolor}{11}{\boxspacing}
\begin{Verbatim}[commandchars=\\\{\}]
\PY{n}{f} \PY{o}{*} \PY{n}{g}
\end{Verbatim}
\end{tcolorbox}

            \begin{tcolorbox}[breakable, size=fbox, boxrule=.5pt, pad at break*=1mm, opacityfill=0]
\prompt{Out}{outcolor}{11}{\boxspacing}
\begin{Verbatim}[commandchars=\\\{\}]
Drinfeld Module morphism:
  From: Drinfeld module defined by T |--> z*t\^{}3 + t\^{}2 + z
  To:   Drinfeld module defined by T |--> (2*z\^{}2 + 4*z + 4)*t\^{}3 + (3*z\^{}2 + 2*z +
2)*t\^{}2 + (2*z\^{}2 + 3*z + 4)*t + z
  Defn: (2*z\^{}2 + 4*z + 4)*t\^{}4 + (z + 1)*t\^{}3 + t\^{}2 + (2*z\^{}2 + 4*z + 4)*t + z
\end{Verbatim}
\end{tcolorbox}
        
    \begin{tcolorbox}[breakable, size=fbox, boxrule=1pt, pad at break*=1mm,colback=cellbackground, colframe=cellborder]
\prompt{In}{incolor}{12}{\boxspacing}
\begin{Verbatim}[commandchars=\\\{\}]
\PY{n}{Frob}\PY{o}{\PYZca{}}\PY{l+m+mi}{5}
\end{Verbatim}
\end{tcolorbox}

            \begin{tcolorbox}[breakable, size=fbox, boxrule=.5pt, pad at break*=1mm, opacityfill=0]
\prompt{Out}{outcolor}{12}{\boxspacing}
\begin{Verbatim}[commandchars=\\\{\}]
Endomorphism of Drinfeld module defined by T |--> z*t\^{}3 + t\^{}2 + z
  Defn: t\^{}15
\end{Verbatim}
\end{tcolorbox}
        
    We observe that the sum of two morphisms \(\phi \to \psi\) is
well-defined. Indeed if \(u\) and \(v\) satisfies the relation
\(\phi_T u = u \psi_T\) and \(\phi_T v = v \psi_T\) then one immediately
deduce that \(\phi_T (u+v) = (u+v) \psi_T\). Therefore
\(\text{Hom}(\phi, \psi)\) is naturally a commutative group for the
addition.

Besides, given that elements of \(\mathbb{F}_q[T]\) define endomorphisms
of any Drinfeld module:
\begin{itemize}
\item the Hom space \(\text{Hom}(\phi, \psi)\)
inherits a structure of left \(\mathbb{F}_q[T]\)-module (composing on
the left by the endomorphism \(\psi_a\) of \(\psi\)),
\item the End space
\(\text{End}(\phi)\) inherits a structure of \(\mathbb{F}_q[T]\)-algebra.
\end{itemize}

Our implementation handles those quite transparently. For example:

    \begin{tcolorbox}[breakable, size=fbox, boxrule=1pt, pad at break*=1mm,colback=cellbackground, colframe=cellborder]
\prompt{In}{incolor}{13}{\boxspacing}
\begin{Verbatim}[commandchars=\\\{\}]
\PY{n}{T} \PY{o}{*} \PY{n}{f}
\end{Verbatim}
\end{tcolorbox}

            \begin{tcolorbox}[breakable, size=fbox, boxrule=.5pt, pad at break*=1mm, opacityfill=0]
\prompt{Out}{outcolor}{13}{\boxspacing}
\begin{Verbatim}[commandchars=\\\{\}]
Drinfeld Module morphism:
  From: Drinfeld module defined by T |--> z*t\^{}3 + t\^{}2 + z
  To:   Drinfeld module defined by T |--> (2*z\^{}2 + 4*z + 4)*t\^{}3 + (3*z\^{}2 + 2*z +
2)*t\^{}2 + (2*z\^{}2 + 3*z + 4)*t + z
  Defn: (2*z\^{}2 + 4*z + 4)*t\^{}4 + (z + 1)*t\^{}3 + t\^{}2 + (2*z\^{}2 + 4*z + 4)*t + z
\end{Verbatim}
\end{tcolorbox}
        
    \begin{tcolorbox}[breakable, size=fbox, boxrule=1pt, pad at break*=1mm,colback=cellbackground, colframe=cellborder]
\prompt{In}{incolor}{14}{\boxspacing}
\begin{Verbatim}[commandchars=\\\{\}]
\PY{n}{Frob} \PY{o}{+} \PY{n}{g}
\end{Verbatim}
\end{tcolorbox}

            \begin{tcolorbox}[breakable, size=fbox, boxrule=.5pt, pad at break*=1mm, opacityfill=0]
\prompt{Out}{outcolor}{14}{\boxspacing}
\begin{Verbatim}[commandchars=\\\{\}]
Endomorphism of Drinfeld module defined by T |--> z*t\^{}3 + t\^{}2 + z
  Defn: (z + 1)*t\^{}3 + t\^{}2 + z
\end{Verbatim}
\end{tcolorbox}
        
    \begin{tcolorbox}[breakable, size=fbox, boxrule=1pt, pad at break*=1mm,colback=cellbackground, colframe=cellborder]
\prompt{In}{incolor}{15}{\boxspacing}
\begin{Verbatim}[commandchars=\\\{\}]
\PY{n}{Frob} \PY{o}{*} \PY{n}{g} \PY{o}{==} \PY{n}{g} \PY{o}{*} \PY{n}{Frob}
\end{Verbatim}
\end{tcolorbox}

            \begin{tcolorbox}[breakable, size=fbox, boxrule=.5pt, pad at break*=1mm, opacityfill=0]
\prompt{Out}{outcolor}{15}{\boxspacing}
\begin{Verbatim}[commandchars=\\\{\}]
True
\end{Verbatim}
\end{tcolorbox}
        
    \begin{tcolorbox}[breakable, size=fbox, boxrule=1pt, pad at break*=1mm,colback=cellbackground, colframe=cellborder]
\prompt{In}{incolor}{16}{\boxspacing}
\begin{Verbatim}[commandchars=\\\{\}]
\PY{p}{(}\PY{n}{Frob} \PY{o}{+} \PY{n}{g}\PY{p}{)}\PY{o}{\PYZca{}}\PY{l+m+mi}{5} \PY{o}{==} \PY{n}{Frob}\PY{o}{\PYZca{}}\PY{l+m+mi}{5} \PY{o}{+} \PY{n}{g}\PY{o}{\PYZca{}}\PY{l+m+mi}{5}
\end{Verbatim}
\end{tcolorbox}

            \begin{tcolorbox}[breakable, size=fbox, boxrule=.5pt, pad at break*=1mm, opacityfill=0]
\prompt{Out}{outcolor}{16}{\boxspacing}
\begin{Verbatim}[commandchars=\\\{\}]
True
\end{Verbatim}
\end{tcolorbox}
        
    \hypertarget{recognizing-isomorphisms}{%
\subsection{Recognizing isomorphisms}\label{recognizing-isomorphisms}}

It follows easily from the definition that a morphism
\(f : \phi \to \psi\) is an isomorphism if and only if the skew
polynomial defining \(f\) is constant.

The method \texttt{is\_isomorphism} allows for checking this fact.

    \begin{tcolorbox}[breakable, size=fbox, boxrule=1pt, pad at break*=1mm,colback=cellbackground, colframe=cellborder]
\prompt{In}{incolor}{17}{\boxspacing}
\begin{Verbatim}[commandchars=\\\{\}]
\PY{n}{g}\PY{o}{.}\PY{n}{is\PYZus{}isomorphism}\PY{p}{(}\PY{p}{)}
\end{Verbatim}
\end{tcolorbox}

            \begin{tcolorbox}[breakable, size=fbox, boxrule=.5pt, pad at break*=1mm, opacityfill=0]
\prompt{Out}{outcolor}{17}{\boxspacing}
\begin{Verbatim}[commandchars=\\\{\}]
False
\end{Verbatim}
\end{tcolorbox}
        
    \begin{tcolorbox}[breakable, size=fbox, boxrule=1pt, pad at break*=1mm,colback=cellbackground, colframe=cellborder]
\prompt{In}{incolor}{18}{\boxspacing}
\begin{Verbatim}[commandchars=\\\{\}]
\PY{n}{h} \PY{o}{=} \PY{n}{phi}\PY{o}{.}\PY{n}{hom}\PY{p}{(}\PY{n}{z}\PY{p}{)}
\PY{n}{h}
\end{Verbatim}
\end{tcolorbox}

            \begin{tcolorbox}[breakable, size=fbox, boxrule=.5pt, pad at break*=1mm, opacityfill=0]
\prompt{Out}{outcolor}{18}{\boxspacing}
\begin{Verbatim}[commandchars=\\\{\}]
Drinfeld Module morphism:
  From: Drinfeld module defined by T |--> z*t\^{}3 + t\^{}2 + z
  To:   Drinfeld module defined by T |--> z*t\^{}3 + (4*z\^{}2 + z + 4)*t\^{}2 + z
  Defn: z
\end{Verbatim}
\end{tcolorbox}
        
    \begin{tcolorbox}[breakable, size=fbox, boxrule=1pt, pad at break*=1mm,colback=cellbackground, colframe=cellborder]
\prompt{In}{incolor}{19}{\boxspacing}
\begin{Verbatim}[commandchars=\\\{\}]
\PY{n}{h}\PY{o}{.}\PY{n}{is\PYZus{}isomorphism}\PY{p}{(}\PY{p}{)}
\end{Verbatim}
\end{tcolorbox}

            \begin{tcolorbox}[breakable, size=fbox, boxrule=.5pt, pad at break*=1mm, opacityfill=0]
\prompt{Out}{outcolor}{19}{\boxspacing}
\begin{Verbatim}[commandchars=\\\{\}]
True
\end{Verbatim}
\end{tcolorbox}
        
    The method \texttt{inverse} computes the inverse of an isomorphism:

    \begin{tcolorbox}[breakable, size=fbox, boxrule=1pt, pad at break*=1mm,colback=cellbackground, colframe=cellborder]
\prompt{In}{incolor}{20}{\boxspacing}
\begin{Verbatim}[commandchars=\\\{\}]
\PY{n}{h}\PY{o}{.}\PY{n}{inverse}\PY{p}{(}\PY{p}{)}
\end{Verbatim}
\end{tcolorbox}

            \begin{tcolorbox}[breakable, size=fbox, boxrule=.5pt, pad at break*=1mm, opacityfill=0]
\prompt{Out}{outcolor}{20}{\boxspacing}
\begin{Verbatim}[commandchars=\\\{\}]
Drinfeld Module morphism:
  From: Drinfeld module defined by T |--> z*t\^{}3 + (4*z\^{}2 + z + 4)*t\^{}2 + z
  To:   Drinfeld module defined by T |--> z*t\^{}3 + t\^{}2 + z
  Defn: 3*z\^{}2 + 4
\end{Verbatim}
\end{tcolorbox}
        
    It is also possible to check if two given Drinfeld modules are
isomorphic using the method \texttt{is\_isomorphic}:

    \begin{tcolorbox}[breakable, size=fbox, boxrule=1pt, pad at break*=1mm,colback=cellbackground, colframe=cellborder]
\prompt{In}{incolor}{21}{\boxspacing}
\begin{Verbatim}[commandchars=\\\{\}]
\PY{n}{phi2} \PY{o}{=} \PY{n}{h}\PY{o}{.}\PY{n}{codomain}\PY{p}{(}\PY{p}{)}
\PY{n}{phi}\PY{o}{.}\PY{n}{is\PYZus{}isomorphic}\PY{p}{(}\PY{n}{phi2}\PY{p}{)}
\end{Verbatim}
\end{tcolorbox}

            \begin{tcolorbox}[breakable, size=fbox, boxrule=.5pt, pad at break*=1mm, opacityfill=0]
\prompt{Out}{outcolor}{21}{\boxspacing}
\begin{Verbatim}[commandchars=\\\{\}]
True
\end{Verbatim}
\end{tcolorbox}
        
    \begin{tcolorbox}[breakable, size=fbox, boxrule=1pt, pad at break*=1mm,colback=cellbackground, colframe=cellborder]
\prompt{In}{incolor}{22}{\boxspacing}
\begin{Verbatim}[commandchars=\\\{\}]
\PY{n}{phi}\PY{o}{.}\PY{n}{is\PYZus{}isomorphic}\PY{p}{(}\PY{n}{psi}\PY{p}{)}
\end{Verbatim}
\end{tcolorbox}

            \begin{tcolorbox}[breakable, size=fbox, boxrule=.5pt, pad at break*=1mm, opacityfill=0]
\prompt{Out}{outcolor}{22}{\boxspacing}
\begin{Verbatim}[commandchars=\\\{\}]
False
\end{Verbatim}
\end{tcolorbox}
        
    In the last case, \(\phi\) and \(\psi\) are isogenous (\emph{via} \(f\))
but they are not isomorphic.

    \hypertarget{the-case-of-finite-fields}{%
\subsection{The case of finite fields}\label{the-case-of-finite-fields}}

When \(K\) is a finite field, the subspace of \(\text{Hom}(\phi, \psi)\)
consisting of morphisms defined by a skew polynomial of degree at most
\(d\) (for some given positive integer \(d\)) is a finite dimensional
\(\mathbb{F}_q\)-vector space.

The method \texttt{basis} on the homset returns a basis of this vector
space:

    \begin{tcolorbox}[breakable, size=fbox, boxrule=1pt, pad at break*=1mm,colback=cellbackground, colframe=cellborder]
\prompt{In}{incolor}{23}{\boxspacing}
\begin{Verbatim}[commandchars=\\\{\}]
\PY{n}{H}\PY{o}{.}\PY{n}{basis}\PY{p}{(}\PY{n}{degree}\PY{o}{=}\PY{l+m+mi}{5}\PY{p}{)}
\end{Verbatim}
\end{tcolorbox}

            \begin{tcolorbox}[breakable, size=fbox, boxrule=.5pt, pad at break*=1mm, opacityfill=0]
\prompt{Out}{outcolor}{23}{\boxspacing}
\begin{Verbatim}[commandchars=\\\{\}]
[Drinfeld Module morphism:
   From: Drinfeld module defined by T |--> z*t\^{}3 + t\^{}2 + z
   To:   Drinfeld module defined by T |--> (2*z\^{}2 + 4*z + 4)*t\^{}3 + (3*z\^{}2 + 2*z
+ 2)*t\^{}2 + (2*z\^{}2 + 3*z + 4)*t + z
   Defn: t + 1,
 Drinfeld Module morphism:
   From: Drinfeld module defined by T |--> z*t\^{}3 + t\^{}2 + z
   To:   Drinfeld module defined by T |--> (2*z\^{}2 + 4*z + 4)*t\^{}3 + (3*z\^{}2 + 2*z
+ 2)*t\^{}2 + (2*z\^{}2 + 3*z + 4)*t + z
   Defn: (2*z\^{}2 + 4*z + 3)*t\^{}4 + z*t\^{}3 + t\^{}2 + (2*z\^{}2 + 4*z + 4)*t + z,
 Drinfeld Module morphism:
   From: Drinfeld module defined by T |--> z*t\^{}3 + t\^{}2 + z
   To:   Drinfeld module defined by T |--> (2*z\^{}2 + 4*z + 4)*t\^{}3 + (3*z\^{}2 + 2*z
+ 2)*t\^{}2 + (2*z\^{}2 + 3*z + 4)*t + z
   Defn: (3*z\^{}2 + 3*z + 1)*t\^{}4 + (3*z\^{}2 + 4*z)*t\^{}3 + (3*z\^{}2 + z + 1)*t\^{}2 + (2*z
+ 3)*t + z\^{}2,
 Drinfeld Module morphism:
   From: Drinfeld module defined by T |--> z*t\^{}3 + t\^{}2 + z
   To:   Drinfeld module defined by T |--> (2*z\^{}2 + 4*z + 4)*t\^{}3 + (3*z\^{}2 + 2*z
+ 2)*t\^{}2 + (2*z\^{}2 + 3*z + 4)*t + z
   Defn: t\^{}4 + t\^{}3]
\end{Verbatim}
\end{tcolorbox}
        
    Over finite fields, we also have one method for checking whether a Hom
space is reduced to zero.

    \begin{tcolorbox}[breakable, size=fbox, boxrule=1pt, pad at break*=1mm,colback=cellbackground, colframe=cellborder]
\prompt{In}{incolor}{24}{\boxspacing}
\begin{Verbatim}[commandchars=\\\{\}]
\PY{n}{H}\PY{o}{.}\PY{n}{is\PYZus{}zero}\PY{p}{(}\PY{p}{)}
\end{Verbatim}
\end{tcolorbox}

            \begin{tcolorbox}[breakable, size=fbox, boxrule=.5pt, pad at break*=1mm, opacityfill=0]
\prompt{Out}{outcolor}{24}{\boxspacing}
\begin{Verbatim}[commandchars=\\\{\}]
False
\end{Verbatim}
\end{tcolorbox}
        
    \begin{tcolorbox}[breakable, size=fbox, boxrule=1pt, pad at break*=1mm,colback=cellbackground, colframe=cellborder]
\prompt{In}{incolor}{25}{\boxspacing}
\begin{Verbatim}[commandchars=\\\{\}]
\PY{n}{psi2} \PY{o}{=} \PY{n}{DrinfeldModule}\PY{p}{(}\PY{n}{A}\PY{p}{,} \PY{p}{[}\PY{n}{z}\PY{p}{,} \PY{l+m+mi}{0}\PY{p}{,} \PY{l+m+mi}{1}\PY{p}{,} \PY{n}{z}\PY{o}{\PYZca{}}\PY{l+m+mi}{2}\PY{p}{]}\PY{p}{)}
\PY{n}{H2} \PY{o}{=} \PY{n}{Hom}\PY{p}{(}\PY{n}{phi}\PY{p}{,} \PY{n}{psi2}\PY{p}{)}
\PY{n}{H2}\PY{o}{.}\PY{n}{is\PYZus{}zero}\PY{p}{(}\PY{p}{)}
\end{Verbatim}
\end{tcolorbox}

            \begin{tcolorbox}[breakable, size=fbox, boxrule=.5pt, pad at break*=1mm, opacityfill=0]
\prompt{Out}{outcolor}{25}{\boxspacing}
\begin{Verbatim}[commandchars=\\\{\}]
True
\end{Verbatim}
\end{tcolorbox}
        
    Besides, when a Hom space is nonzero, we can use the method
\texttt{an\_isogeny} to produce a nonzero morphism in it:

    \begin{tcolorbox}[breakable, size=fbox, boxrule=1pt, pad at break*=1mm,colback=cellbackground, colframe=cellborder]
\prompt{In}{incolor}{26}{\boxspacing}
\begin{Verbatim}[commandchars=\\\{\}]
\PY{n}{H}\PY{o}{.}\PY{n}{an\PYZus{}isogeny}\PY{p}{(}\PY{p}{)}
\end{Verbatim}
\end{tcolorbox}

            \begin{tcolorbox}[breakable, size=fbox, boxrule=.5pt, pad at break*=1mm, opacityfill=0]
\prompt{Out}{outcolor}{26}{\boxspacing}
\begin{Verbatim}[commandchars=\\\{\}]
Drinfeld Module morphism:
  From: Drinfeld module defined by T |--> z*t\^{}3 + t\^{}2 + z
  To:   Drinfeld module defined by T |--> (2*z\^{}2 + 4*z + 4)*t\^{}3 + (3*z\^{}2 + 2*z +
2)*t\^{}2 + (2*z\^{}2 + 3*z + 4)*t + z
  Defn: t + 1
\end{Verbatim}
\end{tcolorbox}
        

\newpage

    \hypertarget{drinfeld-modules-j-invariants}{%
\section{\texorpdfstring{\(j\)-invariants}{j-invariants}}\label{drinfeld-modules-j-invariants}}

    \hypertarget{definition}{%
\subsection{Definition}\label{definition}}

The \(j\)-invariants of a Drinfeld module form a family of elements
which caracterizes the isomorphism class of the Drinfeld module over an
algebraic closure.

More precisely, let \(\phi : \mathbb F_q[T] \to K\{\tau\}\) be a
Drinfeld module and write

\[\phi_T = g_0 + g_1\tau + \cdots + g_r \tau^r\]

where \(r\) denotes the rank of \(\phi\). Let \((k_1, \ldots, k_n)\),
\((d_1, \ldots, d_n)\) be two \(n\)-tuple of integers such that
\(1 \leq k_1 < k_2 < \ldots < k_n < r\) and \(d_i > 0\) for all \(i\).
We consider an additional nonnegative integer \(d\) and assume the
so-called \emph{weight-\(0\) condition}:

\[d_1 (q^{k_1} - 1) + d_2 (q^{k_2} - 1) + \cdots + d_{n} (q^{k_n} - 1) = d (q^r - 1).\]

To this datum, we associate the following quantity, called the
\emph{\(((k_1, \ldots, k_n), (d_1, \ldots, d_n, d))\)-\(j\)-invariant}
of \(\phi\):

\[j_{k_1, \ldots, k_n}^{d_1, \ldots, d_n, d}(\phi) :=
  \frac{1}{g_r^{d}}\prod_{i = 1}^n g_{k_i}^{d_i}\]

A \(j\)-invariant is called \emph{basic} when
\(\gcd(d_1, \ldots, d_n, d) = 1\). There is only a finite number of
basic \(j\)-invariants. Moreover, if
\(((k_1, \ldots, k_n), (d_1, \ldots, d_n, d))\) is any parameter, we can
cook up a basic one simply by dividing the \(d_i\)'s and \(d\) by their
gcd. The \(j\)-invariant corresponding to the initial parameter is equal
to the basic \(j\)-invariant raised to the power
\(\gcd(d_1, \ldots, d_n, d)\).

A classical result, due to Potemine, asserts that two Drinfeld module
\(\phi\) and \(\psi\) are isomorphic over \(\bar K\) if and only if
their basic \(j\)-invariants are all equal.

    \hypertarget{computation-of-j-invariants}{%
\subsection{\texorpdfstring{Computation of
\(j\)-invariants}{Computation of j-invariants}}\label{computation-of-j-invariants}}

    \begin{tcolorbox}[breakable, size=fbox, boxrule=1pt, pad at break*=1mm,colback=cellbackground, colframe=cellborder]
\prompt{In}{incolor}{1}{\boxspacing}
\begin{Verbatim}[commandchars=\\\{\}]
\PY{n}{Fq} \PY{o}{=} \PY{n}{GF}\PY{p}{(}\PY{l+m+mi}{5}\PY{p}{)}
\PY{n}{A}\PY{o}{.}\PY{o}{\PYZlt{}}\PY{n}{T}\PY{o}{\PYZgt{}} \PY{o}{=} \PY{n}{Fq}\PY{p}{[}\PY{p}{]}
\PY{n}{K}\PY{o}{.}\PY{o}{\PYZlt{}}\PY{n}{z}\PY{o}{\PYZgt{}} \PY{o}{=} \PY{n}{Fq}\PY{o}{.}\PY{n}{extension}\PY{p}{(}\PY{l+m+mi}{3}\PY{p}{)}
\PY{n}{phi} \PY{o}{=} \PY{n}{DrinfeldModule}\PY{p}{(}\PY{n}{A}\PY{p}{,} \PY{p}{[}\PY{n}{z}\PY{p}{,} \PY{l+m+mi}{0}\PY{p}{,} \PY{l+m+mi}{1}\PY{p}{,} \PY{n}{z}\PY{p}{,} \PY{n}{z}\PY{o}{\PYZca{}}\PY{l+m+mi}{2}\PY{p}{]}\PY{p}{)}
\end{Verbatim}
\end{tcolorbox}

    The first important thing to know is the (finite) list of parameters
\(((k_1, \ldots, k_n), (d_1, \ldots, d_n, d))\) leading to basic
\(j\)-invariants. This list is accessible \emph{via} the method
\texttt{basic\_j\_invariant\_parameters}.

    \begin{tcolorbox}[breakable, size=fbox, boxrule=1pt, pad at break*=1mm,colback=cellbackground, colframe=cellborder]
\prompt{In}{incolor}{2}{\boxspacing}
\begin{Verbatim}[commandchars=\\\{\}]
\PY{n}{parameters} \PY{o}{=} \PY{n}{phi}\PY{o}{.}\PY{n}{basic\PYZus{}j\PYZus{}invariant\PYZus{}parameters}\PY{p}{(}\PY{p}{)}
\PY{n}{parameters}
\end{Verbatim}
\end{tcolorbox}

            \begin{tcolorbox}[breakable, size=fbox, boxrule=.5pt, pad at break*=1mm, opacityfill=0]
\prompt{Out}{outcolor}{2}{\boxspacing}
\begin{Verbatim}[commandchars=\\\{\}]
[((1, 2, 3), (26, 1, 4, 1)),
 ((1, 2, 3), (57, 1, 3, 1)),
 ((1, 2, 3), (88, 1, 2, 1)),
 ...
 ((1, 2, 3), (93, 26, 153, 32)),
 ((1, 2, 3), (155, 26, 151, 32)),
 ((1, 2, 3), (156, 26, 156, 33))]
\end{Verbatim}
\end{tcolorbox}
        
    For a Drinfeld module of rank \(5\), this list is quite long:

    \begin{tcolorbox}[breakable, size=fbox, boxrule=1pt, pad at break*=1mm,colback=cellbackground, colframe=cellborder]
\prompt{In}{incolor}{3}{\boxspacing}
\begin{Verbatim}[commandchars=\\\{\}]
\PY{n+nb}{len}\PY{p}{(}\PY{n}{parameters}\PY{p}{)}
\end{Verbatim}
\end{tcolorbox}

            \begin{tcolorbox}[breakable, size=fbox, boxrule=.5pt, pad at break*=1mm, opacityfill=0]
\prompt{Out}{outcolor}{3}{\boxspacing}
\begin{Verbatim}[commandchars=\\\{\}]
3402
\end{Verbatim}
\end{tcolorbox}
        
    However, when certain coefficients of \(\phi_T\) vanish, some
\(j\)-invariants will vanish as well as the latter are defined as
products of the formers. In this case, it is useful to know the list of
parameters leading to \emph{nonzero} basic \(j\)-invariants. In order to
get it, we can simply pass in the keyword \texttt{nonzero=True}.

    \begin{tcolorbox}[breakable, size=fbox, boxrule=1pt, pad at break*=1mm,colback=cellbackground, colframe=cellborder]
\prompt{In}{incolor}{4}{\boxspacing}
\begin{Verbatim}[commandchars=\\\{\}]
\PY{n}{nonzero\PYZus{}parameters} \PY{o}{=} \PY{n}{phi}\PY{o}{.}\PY{n}{basic\PYZus{}j\PYZus{}invariant\PYZus{}parameters}\PY{p}{(}\PY{n}{nonzero}\PY{o}{=}\PY{k+kc}{True}\PY{p}{)}
\PY{n}{nonzero\PYZus{}parameters}
\end{Verbatim}
\end{tcolorbox}

            \begin{tcolorbox}[breakable, size=fbox, boxrule=.5pt, pad at break*=1mm, opacityfill=0]
\prompt{Out}{outcolor}{4}{\boxspacing}
\begin{Verbatim}[commandchars=\\\{\}]
[((2, 3), (1, 30, 6)),
 ((2, 3), (6, 24, 5)),
 ((2, 3), (7, 54, 11)),
 ((2, 3), (8, 84, 17)),
 ((2, 3), (9, 114, 23)),
 ((2, 3), (10, 144, 29)),
 ((2, 3), (11, 18, 4)),
 ((2, 3), (13, 78, 16)),
 ((2, 3), (15, 138, 28)),
 ((2, 3), (16, 12, 3)),
 ((2, 3), (17, 42, 9)),
 ((2, 3), (19, 102, 21)),
 ((2, 3), (20, 132, 27)),
 ((2, 3), (21, 6, 2)),
 ((2, 3), (23, 66, 14)),
 ((2, 3), (25, 126, 26))]
\end{Verbatim}
\end{tcolorbox}
        
    Once we know a valid parameter, we can compute the \(j\)-invariant,
using the method \texttt{j\_invariant}:

    \begin{tcolorbox}[breakable, size=fbox, boxrule=1pt, pad at break*=1mm,colback=cellbackground, colframe=cellborder]
\prompt{In}{incolor}{5}{\boxspacing}
\begin{Verbatim}[commandchars=\\\{\}]
\PY{n}{parameter} \PY{o}{=} \PY{n}{nonzero\PYZus{}parameters}\PY{p}{[}\PY{l+m+mi}{0}\PY{p}{]}
\PY{n}{phi}\PY{o}{.}\PY{n}{j\PYZus{}invariant}\PY{p}{(}\PY{n}{parameter}\PY{p}{)}
\end{Verbatim}
\end{tcolorbox}

            \begin{tcolorbox}[breakable, size=fbox, boxrule=.5pt, pad at break*=1mm, opacityfill=0]
\prompt{Out}{outcolor}{5}{\boxspacing}
\begin{Verbatim}[commandchars=\\\{\}]
4*z\^{}2 + 2*z + 1
\end{Verbatim}
\end{tcolorbox}
        
    It is also possible to get the complete list of \(j\)-invariants by
calling the method \texttt{basic\_j\_invariants}:

    \begin{tcolorbox}[breakable, size=fbox, boxrule=1pt, pad at break*=1mm,colback=cellbackground, colframe=cellborder]
\prompt{In}{incolor}{6}{\boxspacing}
\begin{Verbatim}[commandchars=\\\{\}]
\PY{n}{phi}\PY{o}{.}\PY{n}{basic\PYZus{}j\PYZus{}invariants}\PY{p}{(}\PY{n}{nonzero}\PY{o}{=}\PY{k+kc}{True}\PY{p}{)}
\end{Verbatim}
\end{tcolorbox}

            \begin{tcolorbox}[breakable, size=fbox, boxrule=.5pt, pad at break*=1mm, opacityfill=0]
\prompt{Out}{outcolor}{6}{\boxspacing}
\begin{Verbatim}[commandchars=\\\{\}]
\{((2, 3), (1, 30, 6)): 4*z\^{}2 + 2*z + 1,
 ((2, 3), (6, 24, 5)): 4*z + 3,
 ((2, 3), (7, 54, 11)): 2*z,
 ((2, 3), (8, 84, 17)): 4*z\^{}2 + 3*z + 1,
 ((2, 3), (9, 114, 23)): z\^{}2 + 2*z + 1,
 ((2, 3), (10, 144, 29)): 2*z\^{}2 + 2*z + 1,
 ((2, 3), (11, 18, 4)): 2*z + 3,
 ((2, 3), (13, 78, 16)): 3*z\^{}2 + z,
 ((2, 3), (15, 138, 28)): z\^{}2 + 3*z + 1,
 ((2, 3), (16, 12, 3)): 4*z\^{}2 + 3*z + 4,
 ((2, 3), (17, 42, 9)): 3*z\^{}2 + 3*z + 4,
 ((2, 3), (19, 102, 21)): 3*z\^{}2 + 2*z + 4,
 ((2, 3), (20, 132, 27)): 2*z\^{}2 + 2*z + 2,
 ((2, 3), (21, 6, 2)): z\^{}2,
 ((2, 3), (23, 66, 14)): z\^{}2 + 4*z + 1,
 ((2, 3), (25, 126, 26)): 2*z\^{}2 + z + 1\}
\end{Verbatim}
\end{tcolorbox}
        
    \hypertarget{j_k-invariants}{%
\subsubsection{\texorpdfstring{\(j_k\)-invariants}{j\_k-invariants}}\label{j_k-invariants}}

    The \(j\)-invariants associated to the basic parameter

\[\left(k, \left(\frac{q^r-1}{q^{\gcd(r,k)} - 1}, \frac{q^k-1}{q^{\gcd(r,k)} - 1}\right)\right)\]

is known as \(j_k\)-invariants. In this case, the method
\texttt{j\_invariant} accepts the integer \texttt{k} as parameter as
well.

    \begin{tcolorbox}[breakable, size=fbox, boxrule=1pt, pad at break*=1mm,colback=cellbackground, colframe=cellborder]
\prompt{In}{incolor}{7}{\boxspacing}
\begin{Verbatim}[commandchars=\\\{\}]
\PY{n}{phi}\PY{o}{.}\PY{n}{j\PYZus{}invariant}\PY{p}{(}\PY{l+m+mi}{3}\PY{p}{)}
\end{Verbatim}
\end{tcolorbox}

            \begin{tcolorbox}[breakable, size=fbox, boxrule=.5pt, pad at break*=1mm, opacityfill=0]
\prompt{Out}{outcolor}{7}{\boxspacing}
\begin{Verbatim}[commandchars=\\\{\}]
3*z
\end{Verbatim}
\end{tcolorbox}
        
    \begin{tcolorbox}[breakable, size=fbox, boxrule=1pt, pad at break*=1mm,colback=cellbackground, colframe=cellborder]
\prompt{In}{incolor}{8}{\boxspacing}
\begin{Verbatim}[commandchars=\\\{\}]
\PY{n}{phi}\PY{o}{.}\PY{n}{j\PYZus{}invariant}\PY{p}{(}\PY{p}{[}\PY{p}{[}\PY{l+m+mi}{3}\PY{p}{]}\PY{p}{,} \PY{p}{[}\PY{l+m+mi}{156}\PY{p}{,} \PY{l+m+mi}{31}\PY{p}{]}\PY{p}{]}\PY{p}{)}
\end{Verbatim}
\end{tcolorbox}

            \begin{tcolorbox}[breakable, size=fbox, boxrule=.5pt, pad at break*=1mm, opacityfill=0]
\prompt{Out}{outcolor}{8}{\boxspacing}
\begin{Verbatim}[commandchars=\\\{\}]
3*z
\end{Verbatim}
\end{tcolorbox}
        
    \hypertarget{j-invariants-in-small-rank}{%
\subsection{\texorpdfstring{\(j\)-invariants in small
rank}{j-invariants in small rank}}\label{j-invariants-in-small-rank}}

\hypertarget{the-case-of-rank-1}{%
\subsubsection{The case of rank 1}\label{the-case-of-rank-1}}

    \begin{tcolorbox}[breakable, size=fbox, boxrule=1pt, pad at break*=1mm,colback=cellbackground, colframe=cellborder]
\prompt{In}{incolor}{9}{\boxspacing}
\begin{Verbatim}[commandchars=\\\{\}]
\PY{n}{phi} \PY{o}{=} \PY{n}{DrinfeldModule}\PY{p}{(}\PY{n}{A}\PY{p}{,} \PY{p}{[}\PY{n}{z}\PY{p}{,} \PY{l+m+mi}{1}\PY{p}{]}\PY{p}{)}
\PY{n}{phi}
\end{Verbatim}
\end{tcolorbox}

            \begin{tcolorbox}[breakable, size=fbox, boxrule=.5pt, pad at break*=1mm, opacityfill=0]
\prompt{Out}{outcolor}{9}{\boxspacing}
\begin{Verbatim}[commandchars=\\\{\}]
Drinfeld module defined by T |--> t + z
\end{Verbatim}
\end{tcolorbox}
        
    In rank \(1\), there is no basic \(j\)-invariant.

    \begin{tcolorbox}[breakable, size=fbox, boxrule=1pt, pad at break*=1mm,colback=cellbackground, colframe=cellborder]
\prompt{In}{incolor}{10}{\boxspacing}
\begin{Verbatim}[commandchars=\\\{\}]
\PY{n}{phi}\PY{o}{.}\PY{n}{basic\PYZus{}j\PYZus{}invariant\PYZus{}parameters}\PY{p}{(}\PY{p}{)}
\end{Verbatim}
\end{tcolorbox}

            \begin{tcolorbox}[breakable, size=fbox, boxrule=.5pt, pad at break*=1mm, opacityfill=0]
\prompt{Out}{outcolor}{10}{\boxspacing}
\begin{Verbatim}[commandchars=\\\{\}]
[]
\end{Verbatim}
\end{tcolorbox}
        
    It is then not relevant to compute \(j\)-invariant.

Besides, by Potemime's theorem, we conclude that all Drinfeld module of
rank \(1\) with the same defining morphism
\(\gamma : \mathbb F_q[T] \to K\) are isomorphic over an algebraic
closure \(\bar K\) of \(K\). It is not true however that they are
isomorphic over \(K\).

    \begin{tcolorbox}[breakable, size=fbox, boxrule=1pt, pad at break*=1mm,colback=cellbackground, colframe=cellborder]
\prompt{In}{incolor}{11}{\boxspacing}
\begin{Verbatim}[commandchars=\\\{\}]
\PY{n}{psi} \PY{o}{=} \PY{n}{DrinfeldModule}\PY{p}{(}\PY{n}{A}\PY{p}{,} \PY{p}{[}\PY{n}{z}\PY{p}{,} \PY{n}{z}\PY{p}{]}\PY{p}{)}
\PY{n}{phi}\PY{o}{.}\PY{n}{is\PYZus{}isomorphic}\PY{p}{(}\PY{n}{psi}\PY{p}{)}
\end{Verbatim}
\end{tcolorbox}

            \begin{tcolorbox}[breakable, size=fbox, boxrule=.5pt, pad at break*=1mm, opacityfill=0]
\prompt{Out}{outcolor}{11}{\boxspacing}
\begin{Verbatim}[commandchars=\\\{\}]
False
\end{Verbatim}
\end{tcolorbox}
        
    \begin{tcolorbox}[breakable, size=fbox, boxrule=1pt, pad at break*=1mm,colback=cellbackground, colframe=cellborder]
\prompt{In}{incolor}{12}{\boxspacing}
\begin{Verbatim}[commandchars=\\\{\}]
\PY{n}{phi}\PY{o}{.}\PY{n}{is\PYZus{}isomorphic}\PY{p}{(}\PY{n}{psi}\PY{p}{,} \PY{n}{absolutely}\PY{o}{=}\PY{k+kc}{True}\PY{p}{)}
\end{Verbatim}
\end{tcolorbox}

            \begin{tcolorbox}[breakable, size=fbox, boxrule=.5pt, pad at break*=1mm, opacityfill=0]
\prompt{Out}{outcolor}{12}{\boxspacing}
\begin{Verbatim}[commandchars=\\\{\}]
True
\end{Verbatim}
\end{tcolorbox}
        
    \hypertarget{the-case-of-rank-2}{%
\subsubsection{The case of rank 2}\label{the-case-of-rank-2}}

    \begin{tcolorbox}[breakable, size=fbox, boxrule=1pt, pad at break*=1mm,colback=cellbackground, colframe=cellborder]
\prompt{In}{incolor}{13}{\boxspacing}
\begin{Verbatim}[commandchars=\\\{\}]
\PY{n}{phi} \PY{o}{=} \PY{n}{DrinfeldModule}\PY{p}{(}\PY{n}{A}\PY{p}{,} \PY{p}{[}\PY{n}{z}\PY{p}{,} \PY{n}{z}\PY{o}{\PYZca{}}\PY{l+m+mi}{2}\PY{p}{,} \PY{n}{z}\PY{o}{\PYZca{}}\PY{l+m+mi}{3}\PY{p}{]}\PY{p}{)}
\PY{n}{phi}
\end{Verbatim}
\end{tcolorbox}

            \begin{tcolorbox}[breakable, size=fbox, boxrule=.5pt, pad at break*=1mm, opacityfill=0]
\prompt{Out}{outcolor}{13}{\boxspacing}
\begin{Verbatim}[commandchars=\\\{\}]
Drinfeld module defined by T |--> (2*z + 2)*t\^{}2 + z\^{}2*t + z
\end{Verbatim}
\end{tcolorbox}
        
    For Drinfeld module of rank \(2\), there is only one basic
\(j\)-invariant, namely \((1, (q+1, 1))\)

    \begin{tcolorbox}[breakable, size=fbox, boxrule=1pt, pad at break*=1mm,colback=cellbackground, colframe=cellborder]
\prompt{In}{incolor}{14}{\boxspacing}
\begin{Verbatim}[commandchars=\\\{\}]
\PY{n}{phi}\PY{o}{.}\PY{n}{basic\PYZus{}j\PYZus{}invariant\PYZus{}parameters}\PY{p}{(}\PY{p}{)}
\end{Verbatim}
\end{tcolorbox}

            \begin{tcolorbox}[breakable, size=fbox, boxrule=.5pt, pad at break*=1mm, opacityfill=0]
\prompt{Out}{outcolor}{14}{\boxspacing}
\begin{Verbatim}[commandchars=\\\{\}]
[((1,), (6, 1))]
\end{Verbatim}
\end{tcolorbox}
        
    In this case (and only in this case!), it is allowed to call the method
\texttt{j\_invariant} without any parameter.

    \begin{tcolorbox}[breakable, size=fbox, boxrule=1pt, pad at break*=1mm,colback=cellbackground, colframe=cellborder]
\prompt{In}{incolor}{15}{\boxspacing}
\begin{Verbatim}[commandchars=\\\{\}]
\PY{n}{phi}\PY{o}{.}\PY{n}{j\PYZus{}invariant}\PY{p}{(}\PY{p}{)}
\end{Verbatim}
\end{tcolorbox}

            \begin{tcolorbox}[breakable, size=fbox, boxrule=.5pt, pad at break*=1mm, opacityfill=0]
\prompt{Out}{outcolor}{15}{\boxspacing}
\begin{Verbatim}[commandchars=\\\{\}]
4*z\^{}2 + 4
\end{Verbatim}
\end{tcolorbox}
        

\newpage
    
    \hypertarget{drinfeld-modules-norms-and-characteristic-polynomials}{%
\section{\texorpdfstring{Norms and characteristic
polynomials}{Norms and characteristic polynomials}}\label{drinfeld-modules-norms-and-characteristic-polynomials}}

    \hypertarget{definitions}{%
\subsection{Definitions}\label{definitions}}

\hypertarget{the-case-of-endomorphisms}{%
\subsubsection{The case of
endomorphisms}\label{the-case-of-endomorphisms}}

Let \(\phi : \mathbb F_q[T] \to K\{\tau\}\) be a Drinfeld module of
characteristic \(\mathfrak p\) and let \(\mathfrak q\) be a maximal
ideal of \(\mathbb F_q[T]\) different from \(\mathfrak p\).

For each positive integer \(n\), one considers the set of
\(\mathbf q^n\)-torsion points of \(\phi\) defined by:

\[\phi[\mathfrak q^n] = \big\{ x \in \bar K : \phi_a(x) = 0, \forall a \in \mathfrak q^n \big\}\]

where we agree that the variable \(\tau\) in the skew polynomial
\(\phi_a\) acts as the Frobenius \(x \mapsto x^q\). The space
\(\phi[\mathfrak q^n]\) inherits a structure of
\(\mathbb F_q[T]\)-module by letting \(a \in \mathbb F_q[T]\) acting as
\(x \mapsto \phi_a(x)\). By definition \(\phi[\mathfrak q^n]\) is killed
by \(\mathfrak q^n\).

The Tate module of \(\phi\) is defined by:

\[T_{\mathfrak q}(\phi) = \varprojlim_{n \to \infty} \phi[\mathfrak q^n]\]

it is a module over the ring \(A_{\mathfrak q}\), defined as the
completion of \(\mathbb F_q[T]\) at the place \(\mathfrak q\),
\emph{i.e.}
\(A_{\mathfrak q} = \varprojlim_{n \to \infty} \mathbb F_q[T] / \mathfrak q^n\).
If \(r\) denotes the rank of \(\phi\), then it is a standard result that
\(T_{\mathfrak q}(\phi)\) is free of rank \(r\) over
\(A_{\mathfrak q}\).

Any endomorphism \(f : \phi \to \phi\) induces a
\(A_{\mathfrak q}\)-linear map
\(T_{\mathfrak q}(f) : T_{\mathfrak q}(\phi) \to T_{\mathfrak q}(\phi)\).
By definition:
\begin{itemize}
\item the \emph{norm} of \(f\) is the determinant of
\(T_{\mathfrak q}(f)\),
\item the \emph{characteristic polynomial} of \(f\)
is the characteristic polynomial of \(T_{\mathfrak q}(f)\).
\end{itemize}

One proves that they both do not depend on the choice of \(\mathfrak q\)
and that the former lies in \(\mathbb F_q[T]\) while the latter is a
polynomial with coefficients in \(\mathbb F_q[T]\).

\hypertarget{the-case-of-general-morphisms}{%
\subsubsection{The case of general
morphisms}\label{the-case-of-general-morphisms}}

We now consider a morphism \(f : \phi \to \psi\) between two different
Drinfeld modules \(\phi\) and \(\psi\). In this setting, the
characteristic polynomial of \(f\) is no longer defined but the norm of
\(f\) continues to make sense as an ideal (and not an actual element) of
\(\mathbb F_q[T]\).

In order to define it, we consider the application

\[\begin{array}{rcl}
T(f) : \quad \bar K & \to & \bar K \\
x & \mapsto & u(x)
\end{array}\]

where \(u\) is the skew polynomial defining \(f\). We notice that
\(T(f)\) is \(\mathbb F_q[T]\)-linear if we endow the domain and the
codomain with the structure of \(\mathbb F_q[T]\)-module coming from
\(\phi\) and \(\psi\) respectively. If \(f\) is nonzero, the cokernel of
\(T(f)\) is a finitely generated torsion \(\mathbb F_q[T]\)-module;
therefore, it is of the form

\[\text{coker } T(f) \simeq \mathbb F_q[T]/\mathfrak a_1 \times \mathbb F_q[T]/\mathfrak a_2 \times \cdots \times \mathbb F_q[T]/\mathfrak a_m\]

for some ideals \(\mathfrak a_1, \ldots, \mathfrak a_m\) of
\(\mathbb F_q[T]\). The \emph{norm} of \(f\) is defined by
\(\nu(f) = \mathfrak a_1 \cdots \mathfrak a_m\) (product of ideals).

    \hypertarget{computing-norms-and-characteristic-polynomials-in-sagemath}{%
\subsection{Computing norms and characteristic polynomials In
SageMath}\label{computing-norms-and-characteristic-polynomials-in-sagemath}}

As we see, the definition of norms and characteristic polynomials
involve intermediate complicated objects (namely, the Tate module
\(T_{\mathfrak q}(f)\) or the map \(T(f)\)) which themselves involve an
algebraic closure of \(K\) and thus looks difficult to implement while
keeping good performances.

However, our implementation provides facilities for computing
efficiently norms and characteristic polynomials of arbitrary
morphisms/endomorphisms. Those are based on alternative definitions of
norms and characteristic polynomials coming from the theory of Anderson
motives.

    \hypertarget{endomorphisms}{%
\subsubsection{Endomorphisms}\label{endomorphisms}}

    \begin{tcolorbox}[breakable, size=fbox, boxrule=1pt, pad at break*=1mm,colback=cellbackground, colframe=cellborder]
\prompt{In}{incolor}{1}{\boxspacing}
\begin{Verbatim}[commandchars=\\\{\}]
\PY{n}{Fq} \PY{o}{=} \PY{n}{GF}\PY{p}{(}\PY{l+m+mi}{5}\PY{p}{)}
\PY{n}{A}\PY{o}{.}\PY{o}{\PYZlt{}}\PY{n}{T}\PY{o}{\PYZgt{}} \PY{o}{=} \PY{n}{Fq}\PY{p}{[}\PY{p}{]}
\PY{n}{K}\PY{o}{.}\PY{o}{\PYZlt{}}\PY{n}{z}\PY{o}{\PYZgt{}} \PY{o}{=} \PY{n}{Fq}\PY{o}{.}\PY{n}{extension}\PY{p}{(}\PY{l+m+mi}{3}\PY{p}{)}
\PY{n}{phi} \PY{o}{=} \PY{n}{DrinfeldModule}\PY{p}{(}\PY{n}{A}\PY{p}{,} \PY{p}{[}\PY{n}{z}\PY{p}{,} \PY{l+m+mi}{0}\PY{p}{,} \PY{l+m+mi}{1}\PY{p}{,} \PY{n}{z}\PY{p}{]}\PY{p}{)}
\end{Verbatim}
\end{tcolorbox}

    We start our tour by computing the norm of the Frobenius endomorphism:

    \begin{tcolorbox}[breakable, size=fbox, boxrule=1pt, pad at break*=1mm,colback=cellbackground, colframe=cellborder]
\prompt{In}{incolor}{2}{\boxspacing}
\begin{Verbatim}[commandchars=\\\{\}]
\PY{n}{Frob} \PY{o}{=} \PY{n}{phi}\PY{o}{.}\PY{n}{frobenius\PYZus{}endomorphism}\PY{p}{(}\PY{p}{)}
\PY{n}{Frob}\PY{o}{.}\PY{n}{norm}\PY{p}{(}\PY{p}{)}
\end{Verbatim}
\end{tcolorbox}

            \begin{tcolorbox}[breakable, size=fbox, boxrule=.5pt, pad at break*=1mm, opacityfill=0]
\prompt{Out}{outcolor}{2}{\boxspacing}
\begin{Verbatim}[commandchars=\\\{\}]
Principal ideal (T\^{}3 + 3*T + 3) of Univariate Polynomial Ring in T over Finite
Field of size 5
\end{Verbatim}
\end{tcolorbox}
        
    For consistency, the method \texttt{norm} always returns an ideal by
default. This behavior can be overrided by passing in the argument
\texttt{ideal=False}:

    \begin{tcolorbox}[breakable, size=fbox, boxrule=1pt, pad at break*=1mm,colback=cellbackground, colframe=cellborder]
\prompt{In}{incolor}{3}{\boxspacing}
\begin{Verbatim}[commandchars=\\\{\}]
\PY{n}{Frob}\PY{o}{.}\PY{n}{norm}\PY{p}{(}\PY{n}{ideal}\PY{o}{=}\PY{k+kc}{False}\PY{p}{)}
\end{Verbatim}
\end{tcolorbox}

            \begin{tcolorbox}[breakable, size=fbox, boxrule=.5pt, pad at break*=1mm, opacityfill=0]
\prompt{Out}{outcolor}{3}{\boxspacing}
\begin{Verbatim}[commandchars=\\\{\}]
3*T\^{}3 + 4*T + 4
\end{Verbatim}
\end{tcolorbox}
        
    We observe that the norm of the Frobenius is \(\mathbb F_q\)-collinear
to the characteristic, which is a standard fact:

    \begin{tcolorbox}[breakable, size=fbox, boxrule=1pt, pad at break*=1mm,colback=cellbackground, colframe=cellborder]
\prompt{In}{incolor}{4}{\boxspacing}
\begin{Verbatim}[commandchars=\\\{\}]
\PY{n}{phi}\PY{o}{.}\PY{n}{characteristic}\PY{p}{(}\PY{p}{)}
\end{Verbatim}
\end{tcolorbox}

            \begin{tcolorbox}[breakable, size=fbox, boxrule=.5pt, pad at break*=1mm, opacityfill=0]
\prompt{Out}{outcolor}{4}{\boxspacing}
\begin{Verbatim}[commandchars=\\\{\}]
T\^{}3 + 3*T + 3
\end{Verbatim}
\end{tcolorbox}
        
    \begin{tcolorbox}[breakable, size=fbox, boxrule=1pt, pad at break*=1mm,colback=cellbackground, colframe=cellborder]
\prompt{In}{incolor}{5}{\boxspacing}
\begin{Verbatim}[commandchars=\\\{\}]
\PY{l+m+mi}{3} \PY{o}{*} \PY{n}{phi}\PY{o}{.}\PY{n}{characteristic}\PY{p}{(}\PY{p}{)}
\end{Verbatim}
\end{tcolorbox}

            \begin{tcolorbox}[breakable, size=fbox, boxrule=.5pt, pad at break*=1mm, opacityfill=0]
\prompt{Out}{outcolor}{5}{\boxspacing}
\begin{Verbatim}[commandchars=\\\{\}]
3*T\^{}3 + 4*T + 4
\end{Verbatim}
\end{tcolorbox}
        
    We can compute similarly the norm of the endomorphisms \(\phi_a\) for
some \(a \in \mathbb F_q[T]\)

    \begin{tcolorbox}[breakable, size=fbox, boxrule=1pt, pad at break*=1mm,colback=cellbackground, colframe=cellborder]
\prompt{In}{incolor}{6}{\boxspacing}
\begin{Verbatim}[commandchars=\\\{\}]
\PY{n}{f} \PY{o}{=} \PY{n}{phi}\PY{o}{.}\PY{n}{hom}\PY{p}{(}\PY{n}{T}\PY{p}{)}
\PY{n}{f}\PY{o}{.}\PY{n}{norm}\PY{p}{(}\PY{p}{)}
\end{Verbatim}
\end{tcolorbox}

            \begin{tcolorbox}[breakable, size=fbox, boxrule=.5pt, pad at break*=1mm, opacityfill=0]
\prompt{Out}{outcolor}{6}{\boxspacing}
\begin{Verbatim}[commandchars=\\\{\}]
Principal ideal (T\^{}3) of Univariate Polynomial Ring in T over Finite Field of
size 5
\end{Verbatim}
\end{tcolorbox}
        
    \begin{tcolorbox}[breakable, size=fbox, boxrule=1pt, pad at break*=1mm,colback=cellbackground, colframe=cellborder]
\prompt{In}{incolor}{7}{\boxspacing}
\begin{Verbatim}[commandchars=\\\{\}]
\PY{n}{g} \PY{o}{=} \PY{n}{phi}\PY{o}{.}\PY{n}{hom}\PY{p}{(}\PY{n}{T}\PY{o}{+}\PY{l+m+mi}{1}\PY{p}{)}
\PY{n}{g}\PY{o}{.}\PY{n}{norm}\PY{p}{(}\PY{p}{)}
\end{Verbatim}
\end{tcolorbox}

            \begin{tcolorbox}[breakable, size=fbox, boxrule=.5pt, pad at break*=1mm, opacityfill=0]
\prompt{Out}{outcolor}{7}{\boxspacing}
\begin{Verbatim}[commandchars=\\\{\}]
Principal ideal (T\^{}3 + 3*T\^{}2 + 3*T + 1) of Univariate Polynomial Ring in T over
Finite Field of size 5
\end{Verbatim}
\end{tcolorbox}
        
    In each case, we observe that the norm of \(\phi_a\) is the ideal
generated by \(a^3\); it is actually a general fact that the norm of
\(\phi_a\) is \(a^r \mathbb{F}_q[T]\) where \(r\) is the rank of the
underlying Drinfeld module.

    Similarly the characteric polynomial can be computed thanks to the
method \texttt{charpoly} (or \texttt{characteristic\_polynomial}):

    \begin{tcolorbox}[breakable, size=fbox, boxrule=1pt, pad at break*=1mm,colback=cellbackground, colframe=cellborder]
\prompt{In}{incolor}{8}{\boxspacing}
\begin{Verbatim}[commandchars=\\\{\}]
\PY{n}{Frob}\PY{o}{.}\PY{n}{charpoly}\PY{p}{(}\PY{p}{)}
\end{Verbatim}
\end{tcolorbox}

            \begin{tcolorbox}[breakable, size=fbox, boxrule=.5pt, pad at break*=1mm, opacityfill=0]
\prompt{Out}{outcolor}{8}{\boxspacing}
\begin{Verbatim}[commandchars=\\\{\}]
X\^{}3 + (T + 1)*X\^{}2 + (2*T + 3)*X + 2*T\^{}3 + T + 1
\end{Verbatim}
\end{tcolorbox}
        
    \begin{tcolorbox}[breakable, size=fbox, boxrule=1pt, pad at break*=1mm,colback=cellbackground, colframe=cellborder]
\prompt{In}{incolor}{9}{\boxspacing}
\begin{Verbatim}[commandchars=\\\{\}]
\PY{n}{f}\PY{o}{.}\PY{n}{charpoly}\PY{p}{(}\PY{p}{)}
\end{Verbatim}
\end{tcolorbox}

            \begin{tcolorbox}[breakable, size=fbox, boxrule=.5pt, pad at break*=1mm, opacityfill=0]
\prompt{Out}{outcolor}{9}{\boxspacing}
\begin{Verbatim}[commandchars=\\\{\}]
X\^{}3 + 2*T*X\^{}2 + 3*T\^{}2*X + 4*T\^{}3
\end{Verbatim}
\end{tcolorbox}
        
    \begin{tcolorbox}[breakable, size=fbox, boxrule=1pt, pad at break*=1mm,colback=cellbackground, colframe=cellborder]
\prompt{In}{incolor}{10}{\boxspacing}
\begin{Verbatim}[commandchars=\\\{\}]
\PY{n}{g}\PY{o}{.}\PY{n}{charpoly}\PY{p}{(}\PY{n}{var}\PY{o}{=}\PY{l+s+s1}{\PYZsq{}}\PY{l+s+s1}{Y}\PY{l+s+s1}{\PYZsq{}}\PY{p}{)}  \PY{c+c1}{\PYZsh{} We can change the variable name}
\end{Verbatim}
\end{tcolorbox}

            \begin{tcolorbox}[breakable, size=fbox, boxrule=.5pt, pad at break*=1mm, opacityfill=0]
\prompt{Out}{outcolor}{10}{\boxspacing}
\begin{Verbatim}[commandchars=\\\{\}]
Y\^{}3 + (2*T + 2)*Y\^{}2 + (3*T\^{}2 + T + 3)*Y + 4*T\^{}3 + 2*T\^{}2 + 2*T + 4
\end{Verbatim}
\end{tcolorbox}
        
    Again, we can check that the characteristic polynomial of \(\phi_a\) is
\((X-a)^r\):

    \begin{tcolorbox}[breakable, size=fbox, boxrule=1pt, pad at break*=1mm,colback=cellbackground, colframe=cellborder]
\prompt{In}{incolor}{11}{\boxspacing}
\begin{Verbatim}[commandchars=\\\{\}]
\PY{n}{g}\PY{o}{.}\PY{n}{charpoly}\PY{p}{(}\PY{p}{)}\PY{o}{.}\PY{n}{factor}\PY{p}{(}\PY{p}{)}
\end{Verbatim}
\end{tcolorbox}

            \begin{tcolorbox}[breakable, size=fbox, boxrule=.5pt, pad at break*=1mm, opacityfill=0]
\prompt{Out}{outcolor}{11}{\boxspacing}
\begin{Verbatim}[commandchars=\\\{\}]
(X + 4*T + 4)\^{}3
\end{Verbatim}
\end{tcolorbox}
        
    \hypertarget{general-morphisms}{%
\subsubsection{General morphisms}\label{general-morphisms}}

For general morphisms, only the method \texttt{norm} is available (given
that the characteristic polynomial is not defined in this generality).

    \begin{tcolorbox}[breakable, size=fbox, boxrule=1pt, pad at break*=1mm,colback=cellbackground, colframe=cellborder]
\prompt{In}{incolor}{12}{\boxspacing}
\begin{Verbatim}[commandchars=\\\{\}]
\PY{n}{t} \PY{o}{=} \PY{n}{phi}\PY{o}{.}\PY{n}{ore\PYZus{}variable}\PY{p}{(}\PY{p}{)}
\PY{n}{h} \PY{o}{=} \PY{n}{phi}\PY{o}{.}\PY{n}{hom}\PY{p}{(}\PY{n}{t} \PY{o}{+} \PY{l+m+mi}{1}\PY{p}{)}
\PY{n}{h}
\end{Verbatim}
\end{tcolorbox}

            \begin{tcolorbox}[breakable, size=fbox, boxrule=.5pt, pad at break*=1mm, opacityfill=0]
\prompt{Out}{outcolor}{12}{\boxspacing}
\begin{Verbatim}[commandchars=\\\{\}]
Drinfeld Module morphism:
  From: Drinfeld module defined by T |--> z*t\^{}3 + t\^{}2 + z
  To:   Drinfeld module defined by T |--> (2*z\^{}2 + 4*z + 4)*t\^{}3 + (3*z\^{}2 + 2*z +
2)*t\^{}2 + (2*z\^{}2 + 3*z + 4)*t + z
  Defn: t + 1
\end{Verbatim}
\end{tcolorbox}
        
    \begin{tcolorbox}[breakable, size=fbox, boxrule=1pt, pad at break*=1mm,colback=cellbackground, colframe=cellborder]
\prompt{In}{incolor}{13}{\boxspacing}
\begin{Verbatim}[commandchars=\\\{\}]
\PY{n}{h}\PY{o}{.}\PY{n}{norm}\PY{p}{(}\PY{p}{)}
\end{Verbatim}
\end{tcolorbox}

            \begin{tcolorbox}[breakable, size=fbox, boxrule=.5pt, pad at break*=1mm, opacityfill=0]
\prompt{Out}{outcolor}{13}{\boxspacing}
\begin{Verbatim}[commandchars=\\\{\}]
Principal ideal (T + 4) of Univariate Polynomial Ring in T over Finite Field of
size 5
\end{Verbatim}
\end{tcolorbox}
        
    We verify, on examples, that the norm is multiplicative:

    \begin{tcolorbox}[breakable, size=fbox, boxrule=1pt, pad at break*=1mm,colback=cellbackground, colframe=cellborder]
\prompt{In}{incolor}{14}{\boxspacing}
\begin{Verbatim}[commandchars=\\\{\}]
\PY{p}{(}\PY{n}{h} \PY{o}{*} \PY{n}{Frob}\PY{p}{)}\PY{o}{.}\PY{n}{norm}\PY{p}{(}\PY{p}{)} \PY{o}{==} \PY{n}{h}\PY{o}{.}\PY{n}{norm}\PY{p}{(}\PY{p}{)} \PY{o}{*} \PY{n}{Frob}\PY{o}{.}\PY{n}{norm}\PY{p}{(}\PY{p}{)}
\end{Verbatim}
\end{tcolorbox}

            \begin{tcolorbox}[breakable, size=fbox, boxrule=.5pt, pad at break*=1mm, opacityfill=0]
\prompt{Out}{outcolor}{14}{\boxspacing}
\begin{Verbatim}[commandchars=\\\{\}]
True
\end{Verbatim}
\end{tcolorbox}
        
    \begin{tcolorbox}[breakable, size=fbox, boxrule=1pt, pad at break*=1mm,colback=cellbackground, colframe=cellborder]
\prompt{In}{incolor}{15}{\boxspacing}
\begin{Verbatim}[commandchars=\\\{\}]
\PY{p}{(}\PY{n}{h} \PY{o}{*} \PY{n}{f}\PY{p}{)}\PY{o}{.}\PY{n}{norm}\PY{p}{(}\PY{p}{)} \PY{o}{==} \PY{n}{h}\PY{o}{.}\PY{n}{norm}\PY{p}{(}\PY{p}{)} \PY{o}{*} \PY{n}{f}\PY{o}{.}\PY{n}{norm}\PY{p}{(}\PY{p}{)}
\end{Verbatim}
\end{tcolorbox}

            \begin{tcolorbox}[breakable, size=fbox, boxrule=.5pt, pad at break*=1mm, opacityfill=0]
\prompt{Out}{outcolor}{15}{\boxspacing}
\begin{Verbatim}[commandchars=\\\{\}]
True
\end{Verbatim}
\end{tcolorbox}
        
    \begin{tcolorbox}[breakable, size=fbox, boxrule=1pt, pad at break*=1mm,colback=cellbackground, colframe=cellborder]
\prompt{In}{incolor}{16}{\boxspacing}
\begin{Verbatim}[commandchars=\\\{\}]
\PY{p}{(}\PY{n}{Frob} \PY{o}{*} \PY{n}{g}\PY{p}{)}\PY{o}{.}\PY{n}{norm}\PY{p}{(}\PY{p}{)} \PY{o}{==} \PY{n}{Frob}\PY{o}{.}\PY{n}{norm}\PY{p}{(}\PY{p}{)} \PY{o}{*} \PY{n}{g}\PY{o}{.}\PY{n}{norm}\PY{p}{(}\PY{p}{)}
\end{Verbatim}
\end{tcolorbox}

            \begin{tcolorbox}[breakable, size=fbox, boxrule=.5pt, pad at break*=1mm, opacityfill=0]
\prompt{Out}{outcolor}{16}{\boxspacing}
\begin{Verbatim}[commandchars=\\\{\}]
True
\end{Verbatim}
\end{tcolorbox}
        
    \hypertarget{the-case-of-the-frobenius-endomorphism}{%
\subsection{The case of the Frobenius
endomorphism}\label{the-case-of-the-frobenius-endomorphism}}

When \(K\) is a finite field, the characteristic polynomial of the
Frobenius endomorphism is an invariant of primary importance. Our
implementation provides a direct method for accessing it:

    \begin{tcolorbox}[breakable, size=fbox, boxrule=1pt, pad at break*=1mm,colback=cellbackground, colframe=cellborder]
\prompt{In}{incolor}{17}{\boxspacing}
\begin{Verbatim}[commandchars=\\\{\}]
\PY{n}{phi}\PY{o}{.}\PY{n}{frobenius\PYZus{}charpoly}\PY{p}{(}\PY{p}{)}
\end{Verbatim}
\end{tcolorbox}

            \begin{tcolorbox}[breakable, size=fbox, boxrule=.5pt, pad at break*=1mm, opacityfill=0]
\prompt{Out}{outcolor}{17}{\boxspacing}
\begin{Verbatim}[commandchars=\\\{\}]
X\^{}3 + (T + 1)*X\^{}2 + (2*T + 3)*X + 2*T\^{}3 + T + 1
\end{Verbatim}
\end{tcolorbox}
        
    Actually, several different algorithms are available for this
computations. They are accessible by passing in the keyword
\texttt{algorithm} which can be:
\begin{itemize}
\item \texttt{gekeler}: it tries to
identify coefficients by writing that the characteristic polynomial
annihilates the Frobenius endomorphism; this algorithm may fail is some
cases;
\item \texttt{motive}: it uses the action of the Frobenius on the
Anderson motive;
\item \texttt{crystalline}: it uses the action of the
Frobenius on the crystalline cohomology;
\item \texttt{CSA}: it exploits the
structure of central simple algebra of \(\text{Frac }K\{\tau\}\).
\end{itemize}

Of course, all those algorithms output the same answer but performances
may vary. By default, the \texttt{crystalline} algorithm is chosen when
the rank is smaller than the degree of the extension \(K/\mathbb F_q\)
whereas the \texttt{CSA} algorithm is chosen otherwise.

    \hypertarget{isogeny-test}{%
\subsubsection{Isogeny test}\label{isogeny-test}}

An important result asserts that two Drinfeld modules \(\phi\) and
\(\psi\) are isogenous if and only if the characteristic polynomials of
the Frobenius endomorphisms of \(\phi\) and \(\psi\) agrees. This
provides an efficient isogeny test, which is available in our package
\emph{via} the method \texttt{is\_isogenous}.

    \begin{tcolorbox}[breakable, size=fbox, boxrule=1pt, pad at break*=1mm,colback=cellbackground, colframe=cellborder]
\prompt{In}{incolor}{18}{\boxspacing}
\begin{Verbatim}[commandchars=\\\{\}]
\PY{n}{psi} \PY{o}{=} \PY{n}{h}\PY{o}{.}\PY{n}{codomain}\PY{p}{(}\PY{p}{)}  \PY{c+c1}{\PYZsh{} recall that h was an isogeny with domain phi}
\PY{n}{phi}\PY{o}{.}\PY{n}{is\PYZus{}isogenous}\PY{p}{(}\PY{n}{psi}\PY{p}{)}
\end{Verbatim}
\end{tcolorbox}

            \begin{tcolorbox}[breakable, size=fbox, boxrule=.5pt, pad at break*=1mm, opacityfill=0]
\prompt{Out}{outcolor}{18}{\boxspacing}
\begin{Verbatim}[commandchars=\\\{\}]
True
\end{Verbatim}
\end{tcolorbox}
        
    \begin{tcolorbox}[breakable, size=fbox, boxrule=1pt, pad at break*=1mm,colback=cellbackground, colframe=cellborder]
\prompt{In}{incolor}{19}{\boxspacing}
\begin{Verbatim}[commandchars=\\\{\}]
\PY{n}{phi2} \PY{o}{=} \PY{n}{DrinfeldModule}\PY{p}{(}\PY{n}{A}\PY{p}{,} \PY{p}{[}\PY{n}{z}\PY{p}{,} \PY{l+m+mi}{0}\PY{p}{,} \PY{l+m+mi}{1}\PY{p}{,} \PY{n}{z}\PY{o}{\PYZca{}}\PY{l+m+mi}{2}\PY{p}{]}\PY{p}{)}
\PY{n}{phi}\PY{o}{.}\PY{n}{is\PYZus{}isogenous}\PY{p}{(}\PY{n}{phi2}\PY{p}{)}
\end{Verbatim}
\end{tcolorbox}

            \begin{tcolorbox}[breakable, size=fbox, boxrule=.5pt, pad at break*=1mm, opacityfill=0]
\prompt{Out}{outcolor}{19}{\boxspacing}
\begin{Verbatim}[commandchars=\\\{\}]
False
\end{Verbatim}
\end{tcolorbox}
        
    Finally, we mention that, when an isogeny does exist, one can find it
using the method \texttt{an\_isogeny} on the homset.

    \begin{tcolorbox}[breakable, size=fbox, boxrule=1pt, pad at break*=1mm,colback=cellbackground, colframe=cellborder]
\prompt{In}{incolor}{20}{\boxspacing}
\begin{Verbatim}[commandchars=\\\{\}]
\PY{n}{Hom}\PY{p}{(}\PY{n}{phi}\PY{p}{,} \PY{n}{psi}\PY{p}{)}\PY{o}{.}\PY{n}{an\PYZus{}isogeny}\PY{p}{(}\PY{p}{)}
\end{Verbatim}
\end{tcolorbox}

            \begin{tcolorbox}[breakable, size=fbox, boxrule=.5pt, pad at break*=1mm, opacityfill=0]
\prompt{Out}{outcolor}{20}{\boxspacing}
\begin{Verbatim}[commandchars=\\\{\}]
Drinfeld Module morphism:
  From: Drinfeld module defined by T |--> z*t\^{}3 + t\^{}2 + z
  To:   Drinfeld module defined by T |--> (2*z\^{}2 + 4*z + 4)*t\^{}3 + (3*z\^{}2 + 2*z +
2)*t\^{}2 + (2*z\^{}2 + 3*z + 4)*t + z
  Defn: t + 1
\end{Verbatim}
\end{tcolorbox}
        
    \hypertarget{timings}{%
\subsubsection{Timings}\label{timings}}

Below, we compare timings.

When \([K:\mathbb F_q]\) is large, the \texttt{crystalline} algorithm is
the fastest:

    \begin{tcolorbox}[breakable, size=fbox, boxrule=1pt, pad at break*=1mm,colback=cellbackground, colframe=cellborder]
\prompt{In}{incolor}{21}{\boxspacing}
\begin{Verbatim}[commandchars=\\\{\}]
\PY{n}{Fq} \PY{o}{=} \PY{n}{GF}\PY{p}{(}\PY{l+m+mi}{5}\PY{p}{)}
\PY{n}{A}\PY{o}{.}\PY{o}{\PYZlt{}}\PY{n}{T}\PY{o}{\PYZgt{}} \PY{o}{=} \PY{n}{Fq}\PY{p}{[}\PY{p}{]}
\PY{n}{K}\PY{o}{.}\PY{o}{\PYZlt{}}\PY{n}{z}\PY{o}{\PYZgt{}} \PY{o}{=} \PY{n}{Fq}\PY{o}{.}\PY{n}{extension}\PY{p}{(}\PY{l+m+mi}{50}\PY{p}{)}
\PY{n}{phi} \PY{o}{=} \PY{n}{DrinfeldModule}\PY{p}{(}\PY{n}{A}\PY{p}{,} \PY{p}{[}\PY{n}{z}\PY{p}{]} \PY{o}{+} \PY{p}{[}\PY{n}{K}\PY{o}{.}\PY{n}{random\PYZus{}element}\PY{p}{(}\PY{p}{)} \PY{k}{for} \PY{n}{\PYZus{}} \PY{o+ow}{in} \PY{n+nb}{range}\PY{p}{(}\PY{l+m+mi}{5}\PY{p}{)}\PY{p}{]} \PY{o}{+} \PY{p}{[}\PY{l+m+mi}{1}\PY{p}{]}\PY{p}{)}
\end{Verbatim}
\end{tcolorbox}

    \begin{tcolorbox}[breakable, size=fbox, boxrule=1pt, pad at break*=1mm,colback=cellbackground, colframe=cellborder]
\prompt{In}{incolor}{22}{\boxspacing}
\begin{Verbatim}[commandchars=\\\{\}]
\PY{o}{\PYZpc{}}\PY{k}{time} \PYZus{} = phi.frobenius\PYZus{}charpoly(algorithm=\PYZdq{}motive\PYZdq{})
\end{Verbatim}
\end{tcolorbox}

    \begin{Verbatim}[commandchars=\\\{\}]
CPU times: user 969 ms, sys: 5.7 ms, total: 975 ms
Wall time: 981 ms
    \end{Verbatim}

    \begin{tcolorbox}[breakable, size=fbox, boxrule=1pt, pad at break*=1mm,colback=cellbackground, colframe=cellborder]
\prompt{In}{incolor}{23}{\boxspacing}
\begin{Verbatim}[commandchars=\\\{\}]
\PY{o}{\PYZpc{}}\PY{k}{time} \PYZus{} = phi.frobenius\PYZus{}charpoly(algorithm=\PYZdq{}crystalline\PYZdq{})
\end{Verbatim}
\end{tcolorbox}

    \begin{Verbatim}[commandchars=\\\{\}]
CPU times: user 623 ms, sys: 2.98 ms, total: 626 ms
Wall time: 629 ms
    \end{Verbatim}

    \begin{tcolorbox}[breakable, size=fbox, boxrule=1pt, pad at break*=1mm,colback=cellbackground, colframe=cellborder]
\prompt{In}{incolor}{24}{\boxspacing}
\begin{Verbatim}[commandchars=\\\{\}]
\PY{o}{\PYZpc{}}\PY{k}{time} \PYZus{} = phi.frobenius\PYZus{}charpoly(algorithm=\PYZdq{}CSA\PYZdq{})
\end{Verbatim}
\end{tcolorbox}

    \begin{Verbatim}[commandchars=\\\{\}]
CPU times: user 3.92 s, sys: 14.6 ms, total: 3.93 s
Wall time: 3.96 s
    \end{Verbatim}

    On the contrary, when the rank is large, the \texttt{CSA} algorithm is
the fastest:

    \begin{tcolorbox}[breakable, size=fbox, boxrule=1pt, pad at break*=1mm,colback=cellbackground, colframe=cellborder]
\prompt{In}{incolor}{25}{\boxspacing}
\begin{Verbatim}[commandchars=\\\{\}]
\PY{n}{Fq} \PY{o}{=} \PY{n}{GF}\PY{p}{(}\PY{l+m+mi}{5}\PY{p}{)}
\PY{n}{A}\PY{o}{.}\PY{o}{\PYZlt{}}\PY{n}{T}\PY{o}{\PYZgt{}} \PY{o}{=} \PY{n}{Fq}\PY{p}{[}\PY{p}{]}
\PY{n}{K}\PY{o}{.}\PY{o}{\PYZlt{}}\PY{n}{z}\PY{o}{\PYZgt{}} \PY{o}{=} \PY{n}{Fq}\PY{o}{.}\PY{n}{extension}\PY{p}{(}\PY{l+m+mi}{5}\PY{p}{)}
\PY{n}{phi} \PY{o}{=} \PY{n}{DrinfeldModule}\PY{p}{(}\PY{n}{A}\PY{p}{,} \PY{p}{[}\PY{n}{z}\PY{p}{]} \PY{o}{+} \PY{p}{[}\PY{n}{K}\PY{o}{.}\PY{n}{random\PYZus{}element}\PY{p}{(}\PY{p}{)} \PY{k}{for} \PY{n}{\PYZus{}} \PY{o+ow}{in} \PY{n+nb}{range}\PY{p}{(}\PY{l+m+mi}{50}\PY{p}{)}\PY{p}{]} \PY{o}{+} \PY{p}{[}\PY{l+m+mi}{1}\PY{p}{]}\PY{p}{)}
\end{Verbatim}
\end{tcolorbox}

    \begin{tcolorbox}[breakable, size=fbox, boxrule=1pt, pad at break*=1mm,colback=cellbackground, colframe=cellborder]
\prompt{In}{incolor}{26}{\boxspacing}
\begin{Verbatim}[commandchars=\\\{\}]
\PY{o}{\PYZpc{}}\PY{k}{time} \PYZus{} = phi.frobenius\PYZus{}charpoly(algorithm=\PYZdq{}motive\PYZdq{})
\end{Verbatim}
\end{tcolorbox}

    \begin{Verbatim}[commandchars=\\\{\}]
CPU times: user 1.15 s, sys: 12.9 ms, total: 1.16 s
Wall time: 1.17 s
    \end{Verbatim}

    \begin{tcolorbox}[breakable, size=fbox, boxrule=1pt, pad at break*=1mm,colback=cellbackground, colframe=cellborder]
\prompt{In}{incolor}{27}{\boxspacing}
\begin{Verbatim}[commandchars=\\\{\}]
\PY{o}{\PYZpc{}}\PY{k}{time} \PYZus{} = phi.frobenius\PYZus{}charpoly(algorithm=\PYZdq{}crystalline\PYZdq{})
\end{Verbatim}
\end{tcolorbox}

    \begin{Verbatim}[commandchars=\\\{\}]
CPU times: user 2.08 s, sys: 13.9 ms, total: 2.1 s
Wall time: 2.11 s
    \end{Verbatim}

    \begin{tcolorbox}[breakable, size=fbox, boxrule=1pt, pad at break*=1mm,colback=cellbackground, colframe=cellborder]
\prompt{In}{incolor}{28}{\boxspacing}
\begin{Verbatim}[commandchars=\\\{\}]
\PY{o}{\PYZpc{}}\PY{k}{time} \PYZus{} = phi.frobenius\PYZus{}charpoly(algorithm=\PYZdq{}CSA\PYZdq{})
\end{Verbatim}
\end{tcolorbox}

    \begin{Verbatim}[commandchars=\\\{\}]
CPU times: user 40 ms, sys: 1.97 ms, total: 42 ms
Wall time: 41.8 ms
    \end{Verbatim}


\newpage
    
    \hypertarget{drinfeld-modules-exponential-and-logarithm}{%
\section{\texorpdfstring{Exponential and
logarithm}{Exponential and logarithm}}\label{drinfeld-modules-exponential-and-logarithm}}

    \hypertarget{definitions}{%
\subsection{Definitions}\label{definitions}}

Consider the rational function field \(K = \mathbb{F}_q(T)\) and let
\(K_{\infty} = \mathbb{F}_q((\frac 1 T))\) be the completion of this
field at the place \(\frac 1 T\). We define \(\mathbb{C}_{\infty}\) to
be the completion of an algebraic closure of \(K_{\infty}\). The field
\(\mathbb{C}_{\infty}\) plays a similar role to the field of complex
numbers.

An important aspect of Drinfeld modules is their analytic point of view.
Given any Drinfeld module
\(\phi : \mathbb{F}_q[T]\rightarrow\mathbb{C}_{\infty}\{\tau\}\) of rank
\(r\), one can show that there exists a \(\mathbb{F}_q[T]\)-lattice of
rank \(r\), \emph{i.e.} a free \(\mathbb{F}_q[T]\)-submodule of
\(\mathbb{C}_{\infty}\) of rank \(r\), denoted by \(\Lambda_{\phi}\),
such that for any \(z\in \mathbb{C}_{\infty}\) and
\(a\in \mathbb{F}_q[T]\)

\[
  \phi_a(z) = z\prod_{\substack{\lambda\in a^{-1}\Lambda_{\phi}/\Lambda_{\phi} \\ \lambda\neq 0}} \left( 1 - \frac{z}{e_{\phi} (\lambda)} \right)
\]

where \(e_{\phi} : \mathbb{C}_{\infty}\rightarrow\mathbb{C}_{\infty}\)
is a \(\mathbb{F}_q\)-linear, surjective and nonconstant function. In
particular, it may be written as a power series

\[
e_{\phi}(z) = z + \sum_{i\geq 1} \alpha_i z^{q^i}.
\]

This function is called the \emph{exponential} of \(\phi\). The
compositional inverse of \(e_{\phi}\) exists and is named the
\emph{logarithm} of \(\phi\), denoted \(\log_{\phi}\):

\[
\log_{\phi}(z) = z + \sum_{i\geq 1}\beta_i z^{q^i}.
\]

\hypertarget{explanations}{%
\subsection{Explanations}\label{explanations}}

We explain here how the code work. In short, given any Drinfeld module

\[
\phi : T \mapsto g_0 + g_1 \tau + \cdots + g_r \tau^r,
\]

we first compute the logarithm using a recursive procedure and then we
revert the obtained power series to get the exponential.

More precisely, we use the following functional equation:

\[
e_{\phi}(az) = \phi_a(e_{\phi}(z)).
\]

Applying \(\log_{\phi}\) on both side of the functional equation we get

\[
a \log_{\phi}(z) = \log_{\phi}(\phi_a(z)).
\]

Then, one may compare the coefficients on both side of the above
equation and obtain the recursive procedure:

\[
a \beta_i = \sum_{n + m = i} \beta_n g_m^{q^n}.
\]

Setting, \(\beta_0 := 1\), this procedure yields an infinite sequence
\((\beta_i)_{i\geq 0}\) corresponding to the logarithm.

Next, since \(e_{\phi}\circ \log_{\phi}(z) = z\) by definition, we may
now compute the coefficients of \(e_{\phi}\) recursively:

\[
\alpha_i = - \sum_{n = 0}^{i - 1} \alpha_i \beta_{i - n}^{q^n}.
\]

The resulting sequence \((\alpha_i)_{i\geq 0}\) corresponds to the
exponential of \(\phi\).

    \begin{tcolorbox}[breakable, size=fbox, boxrule=1pt, pad at break*=1mm,colback=cellbackground, colframe=cellborder]
\prompt{In}{incolor}{1}{\boxspacing}
\begin{Verbatim}[commandchars=\\\{\}]
\PY{n}{q} \PY{o}{=} \PY{l+m+mi}{4}
\PY{n}{Fq} \PY{o}{=} \PY{n}{GF}\PY{p}{(}\PY{n}{q}\PY{p}{)}
\PY{n}{A} \PY{o}{=} \PY{n}{Fq}\PY{p}{[}\PY{l+s+s1}{\PYZsq{}}\PY{l+s+s1}{T}\PY{l+s+s1}{\PYZsq{}}\PY{p}{]}
\PY{n}{K}\PY{o}{.}\PY{o}{\PYZlt{}}\PY{n}{T}\PY{o}{\PYZgt{}} \PY{o}{=} \PY{n}{Frac}\PY{p}{(}\PY{n}{A}\PY{p}{)}
\PY{n}{phi} \PY{o}{=} \PY{n}{DrinfeldModule}\PY{p}{(}\PY{n}{A}\PY{p}{,} \PY{p}{[}\PY{n}{T}\PY{p}{,} \PY{n}{T}\PY{o}{+}\PY{l+m+mi}{1}\PY{p}{,} \PY{n}{T}\PY{o}{\PYZca{}}\PY{l+m+mi}{2} \PY{o}{\PYZhy{}} \PY{n}{T} \PY{o}{+} \PY{l+m+mi}{1}\PY{p}{]}\PY{p}{)}
\end{Verbatim}
\end{tcolorbox}

    We use the method \texttt{exponential} to compute a power series
approximation of \(e_{\phi}\):

    \begin{tcolorbox}[breakable, size=fbox, boxrule=1pt, pad at break*=1mm,colback=cellbackground, colframe=cellborder]
\prompt{In}{incolor}{2}{\boxspacing}
\begin{Verbatim}[commandchars=\\\{\}]
\PY{n}{exp} \PY{o}{=} \PY{n}{phi}\PY{o}{.}\PY{n}{exponential}\PY{p}{(}\PY{p}{)}
\PY{n}{exp}
\end{Verbatim}
\end{tcolorbox}

            \begin{tcolorbox}[breakable, size=fbox, boxrule=.5pt, pad at break*=1mm, opacityfill=0]
\prompt{Out}{outcolor}{2}{\boxspacing}
\begin{Verbatim}[commandchars=\\\{\}]
z + ((1/(T\^{}3+T\^{}2+T))*z\^{}4) + O(z\^{}8)
\end{Verbatim}
\end{tcolorbox}
        
    To compute the logarithm, use the method \texttt{logarithm}:

    \begin{tcolorbox}[breakable, size=fbox, boxrule=1pt, pad at break*=1mm,colback=cellbackground, colframe=cellborder]
\prompt{In}{incolor}{3}{\boxspacing}
\begin{Verbatim}[commandchars=\\\{\}]
\PY{n}{log} \PY{o}{=} \PY{n}{phi}\PY{o}{.}\PY{n}{exponential}\PY{p}{(}\PY{p}{)}
\PY{n}{log}
\end{Verbatim}
\end{tcolorbox}

            \begin{tcolorbox}[breakable, size=fbox, boxrule=.5pt, pad at break*=1mm, opacityfill=0]
\prompt{Out}{outcolor}{3}{\boxspacing}
\begin{Verbatim}[commandchars=\\\{\}]
z + ((1/(T\^{}3+T\^{}2+T))*z\^{}4) + O(z\^{}8)
\end{Verbatim}
\end{tcolorbox}
        
    The returned power series is a \emph{lazy power series}. This means that
the user does not need to input any precision parameter and the code
computes the coefficients on demands.

    \begin{tcolorbox}[breakable, size=fbox, boxrule=1pt, pad at break*=1mm,colback=cellbackground, colframe=cellborder]
\prompt{In}{incolor}{4}{\boxspacing}
\begin{Verbatim}[commandchars=\\\{\}]
\PY{n}{exp}\PY{p}{[}\PY{n}{q}\PY{o}{*}\PY{o}{*}\PY{l+m+mi}{5}\PY{p}{]}
\end{Verbatim}
\end{tcolorbox}

            \begin{tcolorbox}[breakable, size=fbox, boxrule=.5pt, pad at break*=1mm, opacityfill=0]
\prompt{Out}{outcolor}{4}{\boxspacing}
\begin{Verbatim}[commandchars=\\\{\}]
(T\^{}2045 + T\^{}2044 + ... + T\^{}2 + 1)/(T\^{}4827 + T\^{}4825 + ... + T\^{}342 + T\^{}341)
\end{Verbatim}
\end{tcolorbox}
        
    \begin{tcolorbox}[breakable, size=fbox, boxrule=1pt, pad at break*=1mm,colback=cellbackground, colframe=cellborder]
\prompt{In}{incolor}{5}{\boxspacing}
\begin{Verbatim}[commandchars=\\\{\}]
\PY{n}{log}\PY{p}{[}\PY{n}{q}\PY{o}{*}\PY{o}{*}\PY{l+m+mi}{5}\PY{p}{]}
\end{Verbatim}
\end{tcolorbox}

            \begin{tcolorbox}[breakable, size=fbox, boxrule=.5pt, pad at break*=1mm, opacityfill=0]
\prompt{Out}{outcolor}{5}{\boxspacing}
\begin{Verbatim}[commandchars=\\\{\}]
(T\^{}2045 + T\^{}2044 + ... + T\^{}2 + 1)/(T\^{}4827 + T\^{}4825 + ... + T\^{}342 + T\^{}341)
\end{Verbatim}
\end{tcolorbox}
        
    The logarithm and exponential are the mutual compositional inverse to
one another:

    \begin{tcolorbox}[breakable, size=fbox, boxrule=1pt, pad at break*=1mm,colback=cellbackground, colframe=cellborder]
\prompt{In}{incolor}{6}{\boxspacing}
\begin{Verbatim}[commandchars=\\\{\}]
\PY{n}{log}\PY{o}{.}\PY{n}{compose}\PY{p}{(}\PY{n}{exp}\PY{p}{)}
\end{Verbatim}
\end{tcolorbox}

            \begin{tcolorbox}[breakable, size=fbox, boxrule=.5pt, pad at break*=1mm, opacityfill=0]
\prompt{Out}{outcolor}{6}{\boxspacing}
\begin{Verbatim}[commandchars=\\\{\}]
z + O(z\^{}8)
\end{Verbatim}
\end{tcolorbox}
        
    \begin{tcolorbox}[breakable, size=fbox, boxrule=1pt, pad at break*=1mm,colback=cellbackground, colframe=cellborder]
\prompt{In}{incolor}{7}{\boxspacing}
\begin{Verbatim}[commandchars=\\\{\}]
\PY{n}{exp}\PY{o}{.}\PY{n}{compose}\PY{p}{(}\PY{n}{log}\PY{p}{)}
\end{Verbatim}
\end{tcolorbox}

            \begin{tcolorbox}[breakable, size=fbox, boxrule=.5pt, pad at break*=1mm, opacityfill=0]
\prompt{Out}{outcolor}{7}{\boxspacing}
\begin{Verbatim}[commandchars=\\\{\}]
z + O(z\^{}8)
\end{Verbatim}
\end{tcolorbox}
        
    In order to obtain more coefficients, one can slice the series:

    \begin{tcolorbox}[breakable, size=fbox, boxrule=1pt, pad at break*=1mm,colback=cellbackground, colframe=cellborder]
\prompt{In}{incolor}{8}{\boxspacing}
\begin{Verbatim}[commandchars=\\\{\}]
\PY{n}{exp}\PY{p}{[}\PY{l+m+mi}{0}\PY{p}{:}\PY{n}{q}\PY{o}{\PYZca{}}\PY{l+m+mi}{2} \PY{o}{+} \PY{l+m+mi}{1}\PY{p}{]}  \PY{c+c1}{\PYZsh{} coefficients from 0 to q\PYZca{}2}
\end{Verbatim}
\end{tcolorbox}

            \begin{tcolorbox}[breakable, size=fbox, boxrule=.5pt, pad at break*=1mm, opacityfill=0]
\prompt{Out}{outcolor}{8}{\boxspacing}
\begin{Verbatim}[commandchars=\\\{\}]
[0,
 1,
 0,
 0,
 1/(T\^{}3 + T\^{}2 + T),
 0,
 0,
 0,
 0,
 0,
 0,
 0,
 0,
 0,
 0,
 0,
 (T\^{}14 + T\^{}13 + T\^{}12 + T\^{}10 + T\^{}9 + T\^{}8 + T\^{}6 + T\^{}5 + T\^{}4 + T + 1)/(T\^{}28 + T\^{}24
+ T\^{}20 + T\^{}13 + T\^{}9 + T\^{}5)]
\end{Verbatim}
\end{tcolorbox}
        
    \hypertarget{the-case-of-the-carlitz-module}{%
\subsection{The case of the Carlitz
module}\label{the-case-of-the-carlitz-module}}

Closed formulas for the coefficients of the exponential and logarithm
are known when \(\phi = \rho\), the Carlitz module:

\[
  \rho : T \mapsto T + \tau.
\]

More precisely, for any \(i>0\) we define the following quantities:
\begin{itemize}
\item
\([i] := T^{q^i} - T\),
\item \(D_0 := 1\) and \(D_i := [i][i - 1]^q\cdots [1]^{q^{i-1}}\),
\item \(L_0 := 1\) and \(L_i := [i][i-1]\cdots [1]\).
\end{itemize}

Then, the exponential is given explicitely by: \[
  e_{\rho}(z) = \sum_{i\geq 0} \frac{z^{q^i}}{D_i}
\] and the logarithm is given by \[
  \log_{\rho}(z) = \sum_{i\geq 0} (-1)^i\frac{z^{q^i}}{L_i}.
\]

    \begin{tcolorbox}[breakable, size=fbox, boxrule=1pt, pad at break*=1mm,colback=cellbackground, colframe=cellborder]
\prompt{In}{incolor}{9}{\boxspacing}
\begin{Verbatim}[commandchars=\\\{\}]
\PY{n}{rho} \PY{o}{=} \PY{n}{DrinfeldModule}\PY{p}{(}\PY{n}{A}\PY{p}{,} \PY{p}{[}\PY{n}{T}\PY{p}{,} \PY{l+m+mi}{1}\PY{p}{]}\PY{p}{)}  \PY{c+c1}{\PYZsh{} Carlitz module}
\PY{n}{exp} \PY{o}{=} \PY{n}{rho}\PY{o}{.}\PY{n}{exponential}\PY{p}{(}\PY{p}{)}
\PY{n}{log} \PY{o}{=} \PY{n}{rho}\PY{o}{.}\PY{n}{logarithm}\PY{p}{(}\PY{p}{)}
\end{Verbatim}
\end{tcolorbox}

    \begin{tcolorbox}[breakable, size=fbox, boxrule=1pt, pad at break*=1mm,colback=cellbackground, colframe=cellborder]
\prompt{In}{incolor}{10}{\boxspacing}
\begin{Verbatim}[commandchars=\\\{\}]
\PY{n}{exp}
\end{Verbatim}
\end{tcolorbox}

            \begin{tcolorbox}[breakable, size=fbox, boxrule=.5pt, pad at break*=1mm, opacityfill=0]
\prompt{Out}{outcolor}{10}{\boxspacing}
\begin{Verbatim}[commandchars=\\\{\}]
z + ((1/(T\^{}4+T))*z\^{}4) + O(z\^{}8)
\end{Verbatim}
\end{tcolorbox}
        
    \begin{tcolorbox}[breakable, size=fbox, boxrule=1pt, pad at break*=1mm,colback=cellbackground, colframe=cellborder]
\prompt{In}{incolor}{11}{\boxspacing}
\begin{Verbatim}[commandchars=\\\{\}]
\PY{n}{log}
\end{Verbatim}
\end{tcolorbox}

            \begin{tcolorbox}[breakable, size=fbox, boxrule=.5pt, pad at break*=1mm, opacityfill=0]
\prompt{Out}{outcolor}{11}{\boxspacing}
\begin{Verbatim}[commandchars=\\\{\}]
z + ((1/(T\^{}4+T))*z\^{}4) + O(z\^{}8)
\end{Verbatim}
\end{tcolorbox}
        
    \begin{tcolorbox}[breakable, size=fbox, boxrule=1pt, pad at break*=1mm,colback=cellbackground, colframe=cellborder]
\prompt{In}{incolor}{12}{\boxspacing}
\begin{Verbatim}[commandchars=\\\{\}]
\PY{k}{def} \PY{n+nf}{sq}\PY{p}{(}\PY{n}{i}\PY{p}{)}\PY{p}{:}
    \PY{l+s+sd}{\PYZdq{}\PYZdq{}\PYZdq{}}
\PY{l+s+sd}{    Return [i] = T\PYZca{}(q\PYZca{}i) \PYZhy{} T.}
\PY{l+s+sd}{    \PYZdq{}\PYZdq{}\PYZdq{}}
    \PY{k}{return} \PY{n}{T}\PY{o}{\PYZca{}}\PY{p}{(}\PY{n}{q}\PY{o}{\PYZca{}}\PY{n}{i}\PY{p}{)} \PY{o}{\PYZhy{}} \PY{n}{T}

\PY{k}{def} \PY{n+nf}{D}\PY{p}{(}\PY{n}{i}\PY{p}{)}\PY{p}{:}
    \PY{l+s+sd}{\PYZdq{}\PYZdq{}\PYZdq{}}
\PY{l+s+sd}{    Return 1 if i = 0 and D\PYZus{}i = [i][i \PYZhy{} 1]\PYZca{}q *... * [1]\PYZca{}(q\PYZca{}(i \PYZhy{} 1)) otherwise.}
\PY{l+s+sd}{    \PYZdq{}\PYZdq{}\PYZdq{}}
    \PY{k}{if} \PY{n}{i} \PY{o}{==} \PY{l+m+mi}{0}\PY{p}{:}
        \PY{k}{return} \PY{l+m+mi}{1}
    \PY{k}{return} \PY{n}{prod}\PY{p}{(}\PY{n}{sq}\PY{p}{(}\PY{n}{n}\PY{p}{)}\PY{o}{*}\PY{o}{*}\PY{p}{(}\PY{n}{q}\PY{o}{\PYZca{}}\PY{p}{(}\PY{n}{i} \PY{o}{\PYZhy{}} \PY{n}{n}\PY{p}{)}\PY{p}{)} \PY{k}{for} \PY{n}{n} \PY{o+ow}{in} \PY{n+nb}{range}\PY{p}{(}\PY{l+m+mi}{1}\PY{p}{,} \PY{n}{i} \PY{o}{+} \PY{l+m+mi}{1}\PY{p}{)}\PY{p}{)}

\PY{k}{def} \PY{n+nf}{L}\PY{p}{(}\PY{n}{i}\PY{p}{)}\PY{p}{:}
    \PY{l+s+sd}{\PYZdq{}\PYZdq{}\PYZdq{}}
\PY{l+s+sd}{    Return 1 if i = 0 and L\PYZus{}i = [i][i \PYZhy{} 1] *...* [1] otherwise.}
\PY{l+s+sd}{    \PYZdq{}\PYZdq{}\PYZdq{}}
    \PY{k}{if} \PY{n}{i} \PY{o}{==} \PY{l+m+mi}{0}\PY{p}{:}
        \PY{k}{return} \PY{l+m+mi}{1}
    \PY{k}{return} \PY{n}{prod}\PY{p}{(}\PY{n}{sq}\PY{p}{(}\PY{n}{n}\PY{p}{)} \PY{k}{for} \PY{n}{n} \PY{o+ow}{in} \PY{n+nb}{range}\PY{p}{(}\PY{l+m+mi}{1}\PY{p}{,} \PY{n}{i} \PY{o}{+} \PY{l+m+mi}{1}\PY{p}{)}\PY{p}{)}
\end{Verbatim}
\end{tcolorbox}

    We can check the expected values:

    \begin{tcolorbox}[breakable, size=fbox, boxrule=1pt, pad at break*=1mm,colback=cellbackground, colframe=cellborder]
\prompt{In}{incolor}{13}{\boxspacing}
\begin{Verbatim}[commandchars=\\\{\}]
\PY{k}{assert} \PY{n}{exp}\PY{p}{[}\PY{n}{q}\PY{p}{]} \PY{o}{==} \PY{l+m+mi}{1}\PY{o}{/}\PY{n}{D}\PY{p}{(}\PY{l+m+mi}{1}\PY{p}{)}
\PY{k}{assert} \PY{n}{exp}\PY{p}{[}\PY{n}{q}\PY{o}{*}\PY{o}{*}\PY{l+m+mi}{2}\PY{p}{]} \PY{o}{==} \PY{l+m+mi}{1}\PY{o}{/}\PY{n}{D}\PY{p}{(}\PY{l+m+mi}{2}\PY{p}{)}
\PY{k}{assert} \PY{n}{exp}\PY{p}{[}\PY{n}{q}\PY{o}{*}\PY{o}{*}\PY{l+m+mi}{3}\PY{p}{]} \PY{o}{==} \PY{l+m+mi}{1}\PY{o}{/}\PY{n}{D}\PY{p}{(}\PY{l+m+mi}{3}\PY{p}{)}
\PY{k}{assert} \PY{n}{exp}\PY{p}{[}\PY{n}{q}\PY{o}{*}\PY{o}{*}\PY{l+m+mi}{4}\PY{p}{]} \PY{o}{==} \PY{l+m+mi}{1}\PY{o}{/}\PY{n}{D}\PY{p}{(}\PY{l+m+mi}{4}\PY{p}{)}
\end{Verbatim}
\end{tcolorbox}

    \begin{tcolorbox}[breakable, size=fbox, boxrule=1pt, pad at break*=1mm,colback=cellbackground, colframe=cellborder]
\prompt{In}{incolor}{14}{\boxspacing}
\begin{Verbatim}[commandchars=\\\{\}]
\PY{k}{assert} \PY{n}{log}\PY{p}{[}\PY{n}{q}\PY{p}{]} \PY{o}{==} \PY{o}{\PYZhy{}}\PY{l+m+mi}{1}\PY{o}{/}\PY{n}{L}\PY{p}{(}\PY{l+m+mi}{1}\PY{p}{)}
\PY{k}{assert} \PY{n}{log}\PY{p}{[}\PY{n}{q}\PY{o}{*}\PY{o}{*}\PY{l+m+mi}{2}\PY{p}{]} \PY{o}{==} \PY{l+m+mi}{1}\PY{o}{/}\PY{n}{L}\PY{p}{(}\PY{l+m+mi}{2}\PY{p}{)}
\PY{k}{assert} \PY{n}{log}\PY{p}{[}\PY{n}{q}\PY{o}{*}\PY{o}{*}\PY{l+m+mi}{3}\PY{p}{]} \PY{o}{==} \PY{o}{\PYZhy{}}\PY{l+m+mi}{1}\PY{o}{/}\PY{n}{L}\PY{p}{(}\PY{l+m+mi}{3}\PY{p}{)}
\PY{k}{assert} \PY{n}{log}\PY{p}{[}\PY{n}{q}\PY{o}{*}\PY{o}{*}\PY{l+m+mi}{4}\PY{p}{]} \PY{o}{==} \PY{l+m+mi}{1}\PY{o}{/}\PY{n}{L}\PY{p}{(}\PY{l+m+mi}{4}\PY{p}{)}
\end{Verbatim}
\end{tcolorbox}

    
\end{document}